\documentclass[twocolumn]{aastex62}

\usepackage{graphicx}   
\usepackage{amsmath}    
\usepackage{mathrsfs}    
\usepackage{bm}
\usepackage{amssymb}    
\usepackage{diagbox}    
\usepackage{subfigure}
\usepackage{enumerate}
\usepackage{float}
\usepackage{ulem}
\usepackage{url}
\defcitealias{lizw2019}{Paper I}
\shorttitle{GWR of DDs with ELM WDs}
\shortauthors{Z. Li et al.}

\begin{document}

\title{Gravitational Wave Radiation of Double Degenerates with Extremely low-mass WD companions}

\correspondingauthor{Xuefei Chen}
\email{cxf@ynao.ac.cn}

\author[0000-0002-1421-4427]{Zhenwei Li}
\affiliation{Yunnan Observatories, Chinese Academy of Sciences, Kunming, 650011, People's Republic of China}
\affiliation{Key Laboratory for the Structure and Evolution of Celestial Objects, Chinese Academy of Science, People's Republic of China}
\affiliation{University of the Chinese Academy of Science, Yuquan Road 19, Shijingshan Block, 100049, Beijing, People's Republic of China}


\author{Xuefei Chen}

\affiliation{Yunnan Observatories, Chinese Academy of Sciences, Kunming, 650011, People's Republic of China}
\affiliation{Key Laboratory for the Structure and Evolution of Celestial Objects, Chinese Academy of Science, People's Republic of China}
\affiliation{Center for Astronomical Mega-Science, Chinese Academy of Science, 20A Datun Road, Chaoyang District, Beijing 100012, People's Republic of China} 
\nocollaboration

\author{Hai-Liang Chen}

\affiliation{Yunnan Observatories, Chinese Academy of Sciences, Kunming, 650011, People's Republic of China}
\affiliation{Key Laboratory for the Structure and Evolution of Celestial Objects, Chinese Academy of Science, People's Republic of China}
\nocollaboration

\author{Jiao Li}

\affiliation{Yunnan Observatories, Chinese Academy of Sciences, Kunming, 650011, People's Republic of China}
\affiliation{Key Laboratory for the Structure and Evolution of Celestial Objects, Chinese Academy of Science, People's Republic of China}
\affiliation{University of the Chinese Academy of Science, Yuquan Road 19, Shijingshan Block, 100049, Beijing, People's Republic of China}

\author{Shenghua Yu}
\affiliation{National Astronomical Observatories, Chinese Academy of Sciences, Beijing, 100101, People's Republic of China}

\author{Zhanwen Han}

\affiliation{Yunnan Observatories, Chinese Academy of Sciences, Kunming, 650011, People's Republic of China}
\affiliation{Key Laboratory for the Structure and Evolution of Celestial Objects, Chinese Academy of Science, People's Republic of China}
\affiliation{Center for Astronomical Mega-Science, Chinese Academy of Science, 20A Datun Road, Chaoyang District, Beijing 100012, People's Republic of China} 
\nocollaboration




\begin{abstract}
Double Degenerate systems (DDs) are supposed to be significant gravitational wave (GW) sources for future space-based gravitational-wave detectors, e.g., Laser Interferometer Space Antenna (LISA). Recently, one type of DDs with Extremely low-mass WD (ELM WD; $\lesssim 0.30\; M_\sun$) companions has been largely found in the ELM Survey. They have very short orbital periods and are therefore important sources for LISA detection. Besides, due to the thick envelope of ELM WDs compared with massive WDs (e.g. CO WDs), they are much easier to be found by the combination of electromagnetic (EM) and GW observations. In this paper, we first obtain the population of ELM WDs in DDs with considering the detailed evolutionary tracks of ELM WDs, and then analyse the GW radiation of these systems. We found that about $6\times10^3$ sources could be solely detected by LISA, including $\sim2\times10^3$ chirping sources, and $\sim13$ ($\sim107$) more sources are expected to be detected by both LISA and ELM Survey ({\it{Gaia}}). 
\end{abstract}




\section{Introduction}
\label{sec:1}
Double white dwarfs (DWDs), also known as double degenerates (DDs), are supposed to be dominant sources for 
future space-based gravitational-wave detector \texttt{Laser Interferometer Space Antenna} (LISA, \citealt{lisa2013,lisa2017}). {{Binary population synthesis (BPS) studies show that about several $10^7$ DWDs exist in our Galaxy. Many DWDs will make contributions to the LISA noise (which is called confusion foreground noise), while $(\sim 1-3)\times10^{4}$ sources are expected to be individually detected by LISA}} (\citealt{evans1987,webbink1998,nelemans2001a,liu2009,liu2010,ruiter2010,yus2010,nissanke2012,korol2017,korol2019, lamberts2019}, and references therein). 
{Besides, a growing number of currently operational or future planned optical telescopes, e.g. Large Synoptic Survey Telescope (LSST, \citealt{lsst2009}), {\it{Gaia}} \citep{gaia2014}, etc., will make it possible to detect the electromagnetic (EM) counterparts of gravitational wave (GW) sources \citep{cooray2004,nelemans2006,littenberg2013,korol2017,kupfer2018}.} 
{For example, the recently detected detached DWD ZTF J1539+5027 (ZTF J1539 hereafter) with orbital period of 6.91 minutes, representing the shortest currently known period among detached DWDs, will be identified by LISA with high signal-to-noise ratio (SNR, \citealt{nature2019,littenberg2019}).} 
{The multi-messenger study of DWDs will allow us to get more precise information from these systems 
\citep{shah2014a,shah2014b,littenberg2019}}, which could furthermore give a constraint of the Galaxy structure \citep{korol2019} and shed light on the physics of mass transfer between DWDs \citep{baker2019}. 

In recent years, Brown and his collaborators have found a large number of DWDs with extremely low-mass (ELM, $\lesssim0.30M_\odot$) helium WD companions\footnote{In this paper, we mainly consider ELM WDs with masses less than $0.3M_\odot$, unless otherwise stated. Particularly, we refer to DWDs as binaries consisting of ELM WDs with CO WD companions, which are the most common sources in the ELM Survey.} from the ELM Survey operated on the 6.5m MMT in SAO \citep{brown2010,brown2012,brown2013,brown2016a,kilic2011a,kilic2012,Gianninas2015,brown2020}. 
{Such DWDs have very short orbital periods and the most compact two, J0651+2844 and J0935+4411 (J0651, J0935 hereafter), are expected to be verification sources\footnote{The term ``verification sources'' refers to currently known objects from EM observations, which can be individually detected by \texttt{LISA} \citep{stroeer2006}.} for LISA \citep{brown2011,hermes2012,brown2016a,kupfer2018}.} 
{In comparison to relatively massive WDs, ELM WDs have much thicker H-rich envelopes and longer timescales in proto stage\footnote{The anti-correlation between H-rich envelope mass and the WD mass comes from the competition between radiation pressure and gravity pressure around the time of the birth of a proto (He or CO) WD. For a massive WD, the nuclear reaction rate of H in the envelope, $\dot{M}_{\rm nuc}$, is high due to the high temperature and density at the bottom of th shell, which leads to a large radiation pressure. When the envelope is small enough, the shell-burning energy can be effectively transferred to the surface and radiate away \citep{chen2017}. Then the radiation pressure reduces and can reach equilibrium with the gravity pressure. The timescale of the proto WD stage can be estimated by $M_{\rm env}/\dot{M}_{\rm nuc}$, where $M_{\rm env}$ is the envelope mass of the proto WD. There is a positive correlation between $\dot{M}_{\rm nuc}$ and the core mass of proto WD, while, $M_{\rm env}$ decreases with the proto WD mass. As a consequence, the timescale in the proto stage is longer for lower WD mass.} (after the stripped of envelope and before the extinct of shell burning). {This stage is the most luminous stage except for flashes and the timescales for this phase is proportional to $M_{\rm WD}^{-8}$, where $M_{\rm WD}$ is the mass of proto-He WD} \citep{chen2017}.} 
{In addition, when the WDs enter into the cooling stage, the radii of massive WDs are smaller than that of low-mass WDs with a same effective temperature \citep{romero2019}. All of these make the ELM WDs more likely to be discovered by EM observations.} In fact, most of the detectable detached DWDs with short orbital periods ($<1\;\rm d$) are those with ELM WD companions. The study of DWDs with ELM WDs companions therefore has a significant effect on the future space GW detectors. 

{In general, ELM WDs are mainly produced by binary evolution since the timescale of single stellar evolution to produce He WDs with mass less than $\sim 0.3M_\odot$ is larger than the Universe age \citep{kilic2007}. This theoretical result is consistent with the observations that all of the ELM WDs ($\lesssim0.3M_\odot$) are in binary systems \citep{brown2016a}.} Recently, \citet{lizw2019} have investigated the formation processes of DWDs with ELM WD companions by combining detailed binary evolution calculation with BPS method (\citetalias{lizw2019} hereafter). We show that DWDs with ELM WD companions can be formed from either stable Roche lobe overflow (RL channel) or common envelope (CE) ejection (CE channel). 
{{{The current local space density is expected to be around $1500\;\rm kpc^{-3}$ for a Milky Way-like galaxy assuming a constant star formation rate of $2\;M_\odot\rm yr^{-1}$ }\citep{chomiuk2011}. {This is much larger than the observationally inferred local space density of $160-275 \rm kpc^{-3}$ }\citep{brown2016b,pelisoli2019b}, {and needs to be confirmed by future observations}\footnote{{In \citetalias{lizw2019}, we adopted a simple assumption of constant SFR of 2 $M_\odot\;\rm yr^{-1}$ for our Galaxy. Based on this assumption, the mass of our Galaxy is $2.74\times10^{10}\;M_\odot$ at the age of 13.7 Gyr. This SFR is oversimplified and the past SFR may be larger than 2 $M_\odot\; \rm yr^{-1}$. In order to better compare with observations, a more realistic model is needed.}}.}}
In this paper, with a better Galaxy model, we model the gravitational-wave radiation (GWR) from Galactic DWDs with ELM WDs, investigate their contributions to the confusion foreground noise and study the properties of detectable sources with the combination of EM and GW observations. 

In current work, we only focus on the detached DWDs. Such systems will become semi-detached, and the ELM WDs will become semi-detached at some point and the ELM WDs will transfer material to their companions. Since the mass ratios (the ELM WDs to the companions) is around $\sim \frac 1 4$, the mass transfer process is expected to be stable \citep{han1999}, and these binaries resemble AM CVn stars during the mass transfer \citep{marsh2004}. However, \citet{brown2016b} argued that the majority of He + CO WDs go through unstable mass transfer and merger into single massive ($\sim 1\;M_\odot$) WDs like R CrB stars, by comparing the estimated merger rate of ELM WD binaries with the observed formation rate of AM CVn binaries (See also \citealt{brown2020}). 
The detailed evolution process between ELM WDs and CO WDs will be discussed in the next paper.

This paper is structured as follows. The introduction of GW and EM observations are 
briefly summarized in Section \ref{sec:2}, and the model inputs are given in Section \ref{sec:3}. 
We present the results in Section \ref{sec:4}, including chirp mass distribution, foreground noise, resolved sources, chirping sources, and the corresponding results with the combination of EM 
and GW observations. Uncertainties of our results are presented in Section \ref{sec:5}, Finally, we give the summary and conclusion in Section \ref{sec:6}. 

\section{Observations}
\label{sec:2}
\subsection{GW signals of binary systems}
\label{subsec:2.1}
Only close DWDs ($\lesssim 1\;\rm d$) could contribute to the confusion foreground noise or be resolved as individual sources by LISA. Due to the strong effect of tidal circularization, we then neglect the eccentric effect of GWR by DWDs and assume that all DWDs have circular orbits. For a DWD binary with an orbital period $P_{\rm orb}$, GWR with a frequency of $f_{\rm GW}=2/P_{\rm orb}$, takes away the orbital angular momentum and makes the orbit shrink. The rate of orbital angular momentum loss is written as \citep{landau}
\begin{equation}
  \frac{{\rm d}J_{\rm orb}}{ {\rm d}t}=-\frac{32}{5}\frac{G^{7/2}M_1^2M_2^2\sqrt{M_1+M_2}}{c^5a^{7/2}},
  \label{eq:1}
\end{equation}
where $G$ is the gravitational constant, $c$ is the speed of light in a vacuum, $a$ is the binary separation, and 
$M_1,M_2$ are masses of the two stars, respectively. Inserting the Kepler's third law into Equation (\ref{eq:1}), we have 
\begin{equation}
  \frac{{\rm d}f_{\rm GW}}{ {\rm d}t}=\frac{96}{5}G^{5/3}c^{-5}\pi^{8/3}\mathcal{M}^{5/3}f_{\rm GW}^{11/3},
  \label{eq:2}
\end{equation}
where $\mathcal{M} = (M_1M_2)^{3/5}(M_1+M_2)^{-1/5}$ is the chirp mass. 

{For an inspiralling signal, it is convenient to use a characteristic strain, $h_{\rm c}$, which is given by 
\begin{equation}
  h_c = \sqrt{N_{\rm cycle}}\mathscr{A}
  \label{eq:3}
\end{equation}
for monochromatic sources \citep{finn2000,moore2015}, where ${N_{\rm cycle}}=f_{\rm GW}T_{\rm obs}$, $T_{\rm obs}$ is the integration time of the detectors ($T_{\rm obs} = 4 \rm \;yr$ in this work), and $\mathscr{A}$ is the dimensionless gravitational wave amplitude, i.e. 
\begin{equation}
  \mathscr{A} = \frac{2\pi^{2/3}G^{5/3}f_{\rm GW}^{2/3}\mathcal{M}^{5/3}}{c^4d},
  \label{eq:4}
\end{equation}
where $d$ is the distance of the binary to the Sun.} 

{To estimate the SNR of sources in \textsf{LISA}, we need to calculate the amplitude of the orbit-averaged detector response, $A$, which is expressed as (see equations 42-44 of \citealt{cornish2003}) }
\begin{equation}
  A = \sqrt{|F_+|^2|h_+|^2+|F_\times|^2|h_\times|^2},
  \label{eq:5}
\end{equation}
{where $F_{+}$, $F_{\times}$ are the detector beam patterns, which depend on the sky location and polarization angle of the source (\citealt{cutler1998,cornish2003,rubbo2004,robson2018}, and references therein). $h_+$ and $h_\times$ are the two polarizations of a GW signal emitted by a binary (equation 9-10 in \citealt{korol2017}). Then the SNR of a monochromatic periodic source can be calculated by \citep{moore2015}}
\begin{equation}
  {\rm {SNR}}^2 = \int_0^\infty {\rm d}f\frac{4|\tilde{h}(f)^2|}{S_{\rm n}(f)} = \frac{A^2T_{\rm obs}}{S_{\rm n}(f_{\rm GW})}
  \label{eq:6}
\end{equation}
{where $\tilde{h}(f)$ is the Fourier transform of GW signal as measured by detector, and $S_{\rm n}(f_{\rm GW})$ is the power spectral density of the detector noise at $f_{\rm GW}$. Here we adopt the current configuration for \textsf{LISA} with $2.5\times 10^6\;\rm km$ arm length for the three detector arms, and the total observation time of $4 \;\rm yr$ \citep{lisa2017}. The corresponding sensitivity curve\footnote{\url{github.com/eXtremeGravityInstitute/LISA_Sensitivity}} can be found in \citet{robson2018}. The GWR of unresolved binaries will form the galactic confusion noise, and can be calculated with the updated version of BPS model \citep{nelemans2005,toonen2012} given in \citet{cornish2017}.} Similar to other studies (e.g. \citealt{korol2017,korol2019}), we assume that a source could be detected by LISA if SNR is larger than 7.

\subsection{EM observations}
\label{subsec:2.2}
\subsubsection{ELM Survey}
\label{subsubsec:2.2.1}
{The ELM Survey is a targeted survey of ELM WDs using de-reddened $g-$band magnitudes ($15<g_0<20\;\rm mag$) and color selections to identify targets. It was operated at the 6.5m MMT telescope \citep{brown2010}. {To obtain a high completeness of follow-up observations, }\citet{brown2016a} {defined a ``clean'' sample of ELM WDs in following way:}
\begin{itemize}
\setlength{\itemsep}{0pt}
\setlength{\parsep}{0pt}
\setlength{\parskip}{0pt}
    \item[(1)] Non-variable objects are excluded; the semi-amplitude is restricted by $k>75\;\rm km\;s^{-1}$ and orbital period $P_{\rm orb}$ is less than 2 days based on the sensitivity tests. 
    \item[(2)] Surface gravity is in the range of $4.85<\log g(\rm cm\; s^{-2})<7.15$, to ensure the follow-up observations of ELM Survey are $95\%$ complete.
    \item[(3)] The color selection provides a built-in temperature selection of $8000<T_{\rm eff}<22000\;\rm K$. 
\end{itemize}
Finally there are 62 ELM WDs remaining in the clean sample, where $65\%$ are disk objects and $35\%$ are halo objects from kinematic classification \citep{brown2016b,brown2020}.} 

{The mass distribution of companions of ELM WDs in the clean sample follows a normal distribution with a mean $\mu=0.76M_\odot$ and standard deviation $\sigma = 0.25M_\odot$, which suggests CO WD companions for ELM WDs. The ELM WD binary systems in the clean sample have very short orbital periods ($\lesssim1\;\rm d$) with a median period of $5.4\;\rm hr$ \citep{brown2016a}, and about a half of these systems will merge in 6 Gyr due to the GWR \citep{brown2016b}.} In the following, we will explore whether the DWDs with ELM WDs in the clean sample have detectable GW signals and how many systems could be discovered by both the ELM Survey and GW detectors.

\subsubsection{{\it{Gaia}} data}
\label{subsubsec:2.2.2}
{\it{Gaia}} is a full sky survey with limiting magnitude down to {\it{Gaia}} $m_{\rm G} = 21\;\rm mag$ \citep{GAIA2016}. The {\it{Gaia}} Data Release 2 (DR2, \citealt{GAIADR2_2018}) provides precise astrometric and photometric information for billions of sources, and gives $\sim260000$ high-confidence WD candidates \citep{gentile2019}. From {\it{Gaia}} DR2 data, \citet{pelisoli2019a} find 50 new high-probability (pre-)ELM WDs from 3891 subdwarf A-type stars (sdAs\footnote{Some of the observed properties of sdAs, such as surface gravity and effective temperature, are similar to that of (pre-)ELM WDs. This will lead to some (pre-)ELM WDs being recognized as sdAs \citep{yuj2019}.}) which are previously discovered in the SDSS. This demonstrates that {\it{Gaia}} has the ability to detect more ELM WDs. {Based on the known ELM WDs, evolutionary models and the quality control parameters of \citet{lindegren2018}, \citet{pelisoli2019b} define a color-cut. With this color-cut, they get a catalogue of 5762 ELM WD candidates from {\it{Gaia}} data. More observations are necessary to confirm these candidates and get their stellar parameters, e.g. $\log g, T_{\rm eff}$ (\citealt{widmark2019}).} In this work, we expect to explore the properties of DWDs containing ELM WDs by the combination of LISA and {\it{Gaia}} observations. 

\section{Model inputs and methods}
\label{sec:3}
\subsection{The populations of DWDs with ELM WDs in the Galaxy}
\label{subsec:3.1}
\subsubsection{Parameter space for producing ELM WDs form detailed binary evolution calculation}
\label{subsubsec:3.1.1}
{In this work, we focus on ELM WD binaries with CO WD companions, which are the most common in the ELM Survey. The formation of such ELM WDs has been systematically investigated in \citetalias{lizw2019}. We give a brief summary here. The primary in a progenitor binary (initially more massive one) evolves faster and overfills its Roche lobe on the asymptotic giant branch (AGB) phase. Then the primary transfers mass to the secondary. If the mass ratio is large enough, the mass transfer phase is dynamically unstable and the binary enters into a common envelope (CE) phase. After the ejection of the CE, a CO WD + MS binary is formed. The secondary (The MS companion here) evolves further and fills its Roche lobe at some point.} If the secondary starts mass transfer in its late MS or during the Hertzsprung gap (HG) phase, the mass transfer is dynamically stable and a (pre-)ELM WD is produced at the end of mass transfer. This formation channel is called RL channel. On the contrary, if the secondary starts mass transfer during HG or near the base of the red giant branch (RGB), the mass transfer can be dynamically unstable. Then the system will enter a CE phase. ELM WD will be produced after the ejection of the CE, and this formation channel is CE channel (See Figure 2 in \citetalias{lizw2019}). 

{{To get the number/birthrate of DWDs with ELM WD components in the Galaxy, we should first know which binary can produce such objects. We then did comprehensive binary evolution calculations, using the state-of-the-art stellar evolution code Modules for Experiments in Stellar Astrophysics (\texttt{MESA}, version 9575, }\citealt{paxton2011,paxton2013,paxton2015}{), to obtain the parameter space for producing DWDs with ELM WD components from RL channel.}}

{{Our detailed binary evolution calculations start from binaries consisting of a CO WD and a zero-age MS star. The mass of the CO WD ranges from 0.45 to 1.1 $M_\odot$, and the MS ranges from 0.8 to $2.0\; M_\odot$.} {The initial orbital period ranges from a minimum period for which the MS star fills its Roche lobe at zero-age main sequence, and then continues to an upper limit of the orbital period for which the mass transfer rate is up to $10^{-4}M_\odot\rm yr^{-1}$ (the mass transfer is considered to be dynamically unstable in our calculation) or the He core mass is larger than $0.3\;M_\odot$.} We stop our calculation if the evolutionary time is larger than 13.7 Gyr.}
{Such binary evolution calculation for Population I stars with a metallicity of $Z=0.02$ has been done in \citetalias{lizw2019}. In the calculation, the mass accumulation efficiency of CO WDs is computed with following formula:} 
\begin{equation}
    \dot{M}_{\rm CO} = \eta_{\rm H}\eta_{\rm He}|\dot{M}_{\rm d}|
    \label{eq:7}
\end{equation}
where $\eta_{\rm H}$ and $\eta_{\rm He}$ are the mass accumulation efficiency for hydrogen burning and helium burning, respectively. We adopt the description of \citet{hachisu1999} and \citet{kato2004} for $\eta_{\rm H}$ and $\eta_{\rm He}$, respectively.

{In order to model the evolution of binaries in the halo, we did the similar binary evolution calculations for $Z=0.001$ in this paper. The hydrogen mass fraction $X$ is computed as $X=0.76-3Z$ \citep{pols1998}. Here we simply assume that mass accumulation efficiency for $Z=0.001$ are same as $Z=0.2$ (see discussion in \citealt{meng2009,chenh2019}).}

\begin{figure}
    \centering
    \includegraphics[width=\columnwidth]{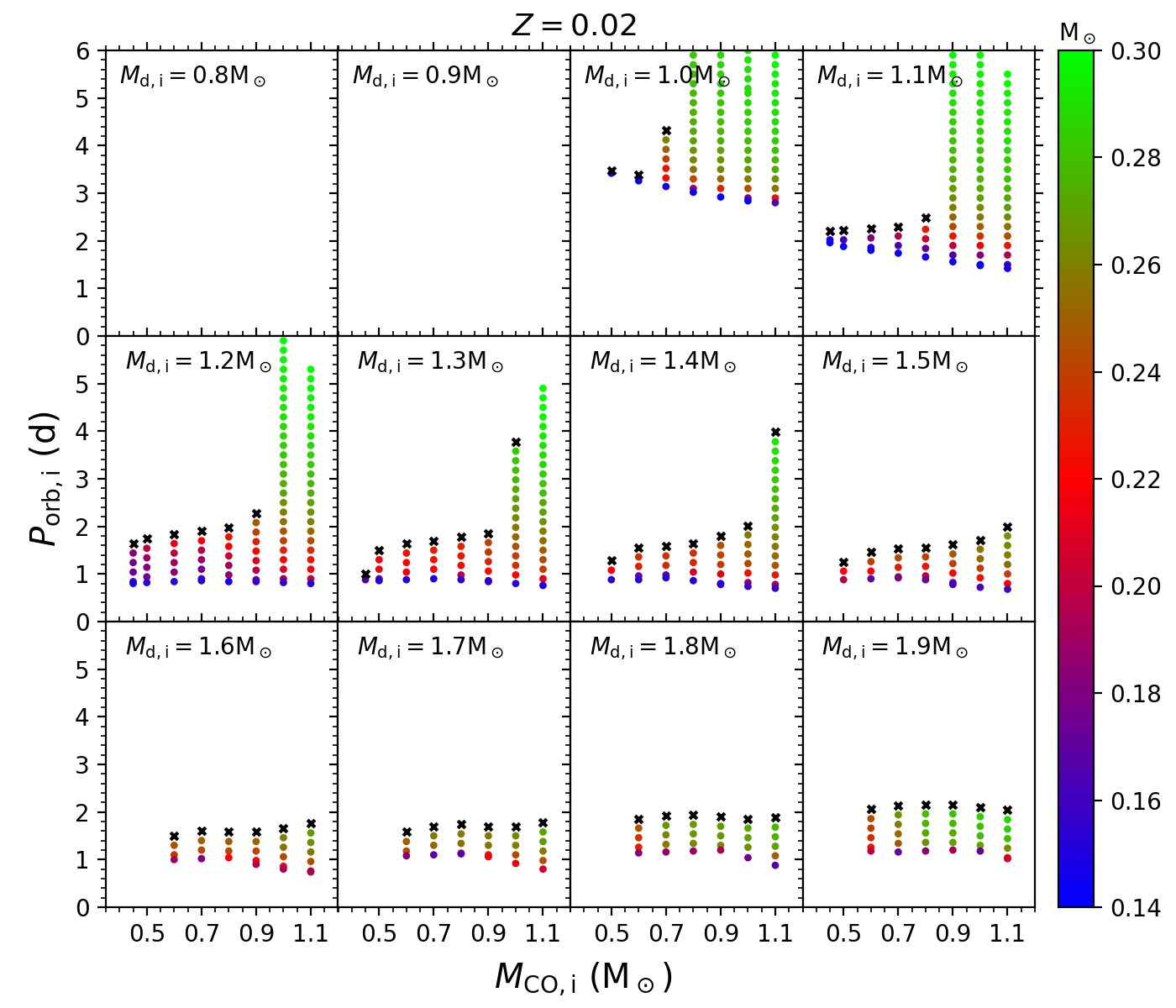}
    \includegraphics[width=\columnwidth]{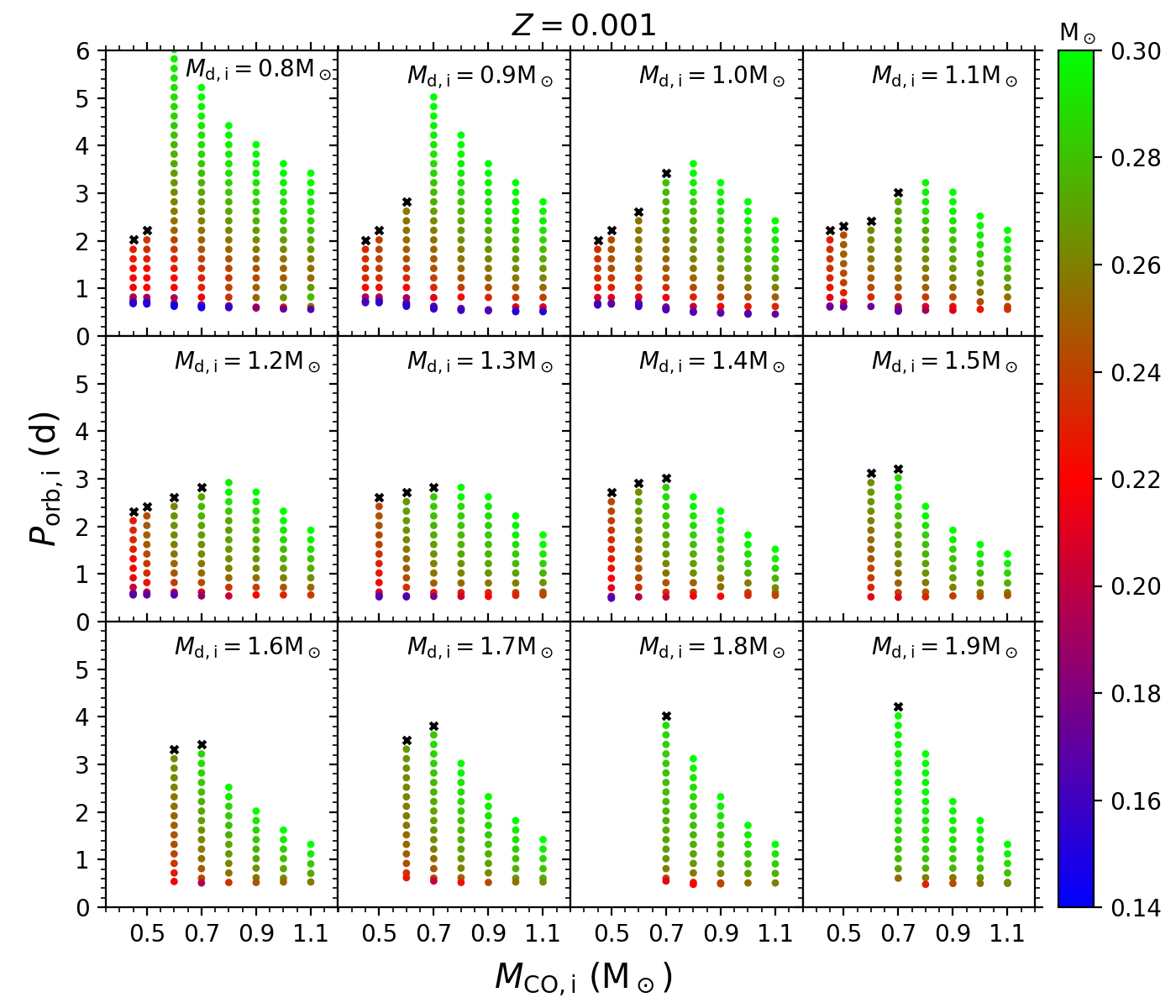}
    \caption{The parameter space for producing ELM WDs from RL channel of $Z = 0.02, 0.001$ in the CO WD mass–initial orbital period plane ($M_{\rm CO,i}-P_{\rm orb,i}$ plane). The initial donor mass is indicated in each panel. {The parameter grids of $M_{\rm d,i}=2.0\;M_\odot$ are not shown in these two cases.} The final masses of ELM WDs are indicated with color bar. The minimum orbital period is determined by bifurcation period and the upper boundary of orbital period is determined by the maximum mass transfer rate of $10^{-4}\;M_\odot \rm yr^{-1}$, as shown in black crosses, or the maximum ELM WD mass of $0.3\;M_\odot$.}
    \label{fig:1}
\end{figure}

The parameter spaces for $Z=0.02$ and $Z=0.001$ are shown in Figure \ref{fig:1}, where $M_{\rm CO,i}$ and $P_{\rm orb,i}$ are the initial CO WD mass and initial orbital period, respectively. The initial donor mass, $M_{\rm d,i}$, is indicated in each panel. 
{{The minimum orbital period is determined by the bifurcation period in low-mass binary evolution. If the initial orbital period is shorter than the bifurcation period, the donor star cannot develop a compact He core in its life and evolve to a brown dwarf likely} \citep{chen2017}.} The upper boundary of orbital period is determined by the maximum mass transfer rate of $10^{-4}\;M_\odot \rm yr^{-1}$ (marked with black crosses), or the maximum helium WDs of $0.3\;M_\odot$. 
{{For these plots, we can find that the parameter space for $Z=0.001$ is obviously larger than that for $Z=0.02$ for given MS mass. Furthermore, due to the shorter lifetimes of the low-$Z$ stars, the MS stars with mass as low as $0.8\;M_\odot$ can also contribute to the production of ELM WDs. As a consequence, DWDs with ELM WDs are more likely produced from the RL channel in low-$Z$ environment than that in the high $Z$ environment as shown in Section \ref{subsec:4.1}.}} 

\subsubsection{The populations of ELM WDs in the Galaxy from BPS}
\label{subsubsec:3.1.2}

To obtain the population of DWDs with ELM WDs in the Galaxy, we first need to produce CO WD binaries from BPS simulations for $Z=0.02$ (for the bulge and the disk) and $Z=0.001$ (for the halo), respectively. In the BPS, we need to generate primordial binaries. The main input distributions are introduced below.
\begin{itemize}
\setlength{\itemsep}{0pt}
\setlength{\parsep}{0pt}
\setlength{\parskip}{0pt}
    \item[(1)] Star formation rate (SFR). Similar to \citet{yus2010}, we assume a quasi-exponential SFR for the bulge and the disk, i.e., 
\begin{equation}
  {\rm SFR}(t) =
    \begin{cases}
	10.6\exp^{-(t-t_0)/\tau} \\
	\quad+ 0.125(t-t_0) {\rm M_\odot yr^{-1}}, \qquad t>t_0 \\
    0, \qquad\qquad\qquad\qquad\qquad\quad 0<t<t_0
    \end{cases}
    \label{eq:8}
\end{equation}
{Assuming the current age of the Galaxy 13.7 Gyr, $t_0 = 4\;\rm Gyr$ gives an age of 9.7 Gyr for the bulge and disk. And $\tau=9\;\rm Gyr$ yields an SFR of 4.82 $M_\odot\;\rm yr^{-1}$ for the Galaxy, with $1.45\;M_\sun\rm yr^{-1}$ for the bulge and $3.37\;M_\sun\rm yr^{-1}$ for the disc, which are consistent with the observations \citep{diehl2006,fantin2019}.} 

Regarding the star formation history of the halo, a single starburst with total mass of $0.7\times 10^{9}$ at $t = 0$ is assumed for the halo \citep{bland2016}. Then we obtain the total mass of the Galaxy, i.e., $M_{\rm Gal}\simeq 6.9\times10^{10}\;\rm M_\odot$, and the mass of the bulge, disc, and halo are $2\times10^{10}\;\rm M_\odot$, $4.8\times10^{10}\;\rm M_\odot$ and $0.7\times10^9\;\rm M_\odot$, respectively (See more details in Section \ref{subsec:3.2}).
    \item[(2)] Initial mass function IMF. The primary mass is given by the following IMF\footnote{{We have checked the influence of different IMFs on our results and find that about $10\%$ less number of DWDs with ELM WDs are produced if we adopted IMF from} \citep{kroupa2001}.} \citep{miller1979,eggleton1989}:
\begin{equation}
   M = \frac{0.19X'}{(1-X')^{0.75}+0.032(1-X')^{0.25}},
   \label{eq:9}
\end{equation}
where $X'$ is a random number between 0 and 1, which gives the mass ranging from 0.1 to 100 $\rm M_\odot$.
    \item[(3)] Initial mass ratio distribution \citep{mazeh1992}: 
\begin{equation}
  n(q')=1,\; q' \lesssim 1
  \label{eq:10}
\end{equation}
where $q'$ represents the mass ratio of initial binary. 
    \item[(4)] Distribution of initial separation \citep{han1998}: 
\begin{equation}
    an(a)=
    \begin{cases}
    0.07(a/a_0)^{1.2},\qquad a\le a_0 \\
    0.07, \qquad \qquad \quad a_0 \le a \le a_1,
    \end{cases}
    \label{eq:11}
\end{equation}
where $a_0=10\;\rm{R}_{\odot}$, $a_1=5.75\times 10^6\;\rm{R}_{\odot}=0.13\;\rm pc$. {This distribution implies that approximately 50 percent of systems are binary systems with orbital periods less than 100 yrs, i.e., the initial binary fraction is assumed to be 50 percent (see also \citealt{han1995}).} 
\end{itemize}

For each BPS simulation, we generate $5\times 10^6$ primordial binaries (with $Z=0.02$ for the disk and bulge, $Z=0.001$ for the halo) through Monte-Carlo method, then evolve these binaries using the rapid binary-star evolution code \texttt{BSE} \citep{hurley00,hurley02} and obtain a number of CO WD binaries in which the companions are just filling their Roche lobes and begin to transfer material to the CO WDs. 

In the next step, we interpolate the parameters of CO WD binaries into the corresponding binary evolution grid ($Z=0.02,0.001$) and obtain the populations of the DWDs with ELM WDs from the RL channel. 

If a binary has parameters outside of the grid spaces, it either cannot produce ELM WD or enters the CE evolution phase. The latter can also produce DWDs with ELM WDs as shown in \citetalias{lizw2019}. {We use the standard energy budget prescription to treat this process in \texttt{BSE}. In this prescription, a part of released orbital energy during the spiral-in process is used to eject the envelope\footnote{{To explain some observed DWDs,} \citet{nelemans2000} {suggested an alternative CE ejection mechanism, named $\gamma-$ formalism. In this scenario, the orbital angular momentum is carried away by the mass loss. However, the physical explanation of $\gamma-$ formalism is still unclear. Therefore, in this work, we adopt the widely accepted standard energy budget prescription to treat the CE process.}} \citep{webbink1984,livio1988,dekool1990}. {Two parameters, i.e. the CE ejection efficiency $\alpha_{\rm CE}$ and the dimensionless structure parameter $\lambda$, are introduced for the energy budget prescription. In }\citetalias{lizw2019}, {we adopted three values of the combined parameter $\alpha_{\rm CE}\lambda$ with 0.25, 0.5, 1, respectively.} {{We find that, the model with $\alpha_{\rm CE}\lambda=1$ can well reproduce the mass peak ($\sim 0.18 M_\odot$) of the ELM WDs and the mass distribution for the CO WD companions. But the predicted local space density is much higher than that of observations as mentioned in Section \ref{sec:1}, indicating that a larger number DWDs with ELM WDs are not found yet.}} In this paper, we adopt $\alpha_{\rm CE}\lambda=1$ and the effect of uncertainties of these parameters on our final results is discussed in Section \ref{subsec:5.1}.}

Finally, with BPS model and the star formation history of the Galaxy, we can simulate the populations of DDs with ELM WDs in the Galaxy. 

\subsection{The Galaxy model}
\label{subsec:3.2}
The GW strains of binary systems are inversely proportional to the distance. To obtain the distance of each simulated system to the Sun, it is crucial to model the space distribution of stars in the Galaxy. However, the Galaxy structure is very complicated due to the uncertainty of its star formation history, dynamics, and chemical evolution, etc (see the most recent review of \citealt{bland2016}). For convenience, here we adopt the spherical power-law model of bulge, two-component model for the disk \citep{binney2008} and an isotropic double-power-law model for the halo \citep{kafle2014}. {As introduced in Section \ref{subsubsec:3.1.2}, the age of halo (bulge and disk) is assumed to be $13.7 \;\rm Gyr$ (9.7 Gyr).} The influence of the uncertainties of the Galaxy model on our results is discussed in Section \ref{subsec:5.2}. 

The spatial density of the Galaxy, $\rho_{\rm Gal}$, is written as 
\begin{equation}
  \rho_{\rm Gal}=\rho_{\rm B}+\rho_{\rm D}+\rho_{\rm H},
    \label{eq:12}
\end{equation}
where $\rho_{\rm B}$, $\rho_{\rm D}$, and $\rho_{\rm H}$ are the spatial density of the bulge, disk and halo, respectively. 
The density of bulge is expressed as  
\begin{equation}
	\rho_{\rm B}(R,z) = \rho_{\rm B,0}\times
    \begin{cases}
	  \left(\frac{q}{a_{\rm b}}\right)^{-\alpha_{\rm b}} \exp\left(-\frac{q^2}{r_{\rm b}^2}\right), \quad q>q_{\rm min}, \\
	  \left(\frac{q_{\rm min}}{a_{\rm b}}\right)^{-\alpha_{\rm b}} \exp\left(-\frac{q^2}{r_{\rm b}^2}\right), {\rm{otherwise}}, \\
    \end{cases}
    \label{eq:13}
\end{equation}
where $(R,z)$ is the Galactocentric distance and height in cylindrical coordinates, and 
$q = (R^2+\frac{z^2}{q_{\rm b}^2})^{1/2}$, and $\rho_{\rm B,0} (= 1.722 \;M_\sun \;\rm pc^{-3})$ is 
the central mass density of the bulge. The cutoff with $q_{\rm min}=10^{-2}\;\rm kpc$ is a modification to prevent infinite density at the Galactic center \citep{galaxy2016}. 
Other values of parameters in the bulge are taken from \citet{binney2008}, i.e., 
$\alpha_{\rm b}=1.8$, $a_{\rm b}=1.0\;\rm kpc$, 
$r_{\rm b}=1.9\;\rm kpc$, and $q_{\rm b}=0.6$. 

For the disk, we have 
\begin{eqnarray}
  \rho_{\rm D}(R,z)&=&\frac{\sum_{\rm t}}{2z_{\rm t}}\exp\left(-\frac{R}{R_{\rm t}}-\frac{|z|}{z_{\rm t}}\right) \nonumber\\
					&+&\frac{\sum_{\rm T}}{2z_{\rm T}}\exp\left(-\frac{R}{R_{\rm T}}-\frac{|z|}{z_{\rm T}}\right).
    \label{eq:14}
\end{eqnarray}
where $\sum_{\rm t}=970.294\;M_\sun\;\rm pc^{-2}$ and $\sum_{\rm T}=268.648\;M_\sun\;\rm pc^{-2}$ are the central surface densities of the thin and thick disk, respectively. 
Other values of parameters in the disc are taken from \citet{bland2016}, i.e., 
$R_{\rm t}=2.6\;\rm kpc$, $R_{\rm T}=2\;\rm kpc$, 
$z_{\rm t}=0.3\;\rm kpc$, and $z_{\rm T}=0.9\;\rm kpc$. 

The density of stellar halo is written as 
\citep{kafle2014}, 
\begin{equation}
  \rho_{\rm H}(r_{\rm G}) = \rho_{\rm H,0}\times
    \begin{cases}
	  \left(\frac{r_{\rm h,min}}{r_{\rm h,b}}\right)^{-\alpha_{\rm h,1}}, r_{\rm G}<r_{\rm h,min}, \\
	  \left(\frac{r_{\rm G}}{r_{\rm h,b}}\right)^{-\alpha_{\rm h,1}}, r_{\rm h,min}\leq r_{\rm G}<r_{\rm h,b}, \\
	  \left(\frac{r_{\rm h,G}}{r_{\rm h,b}}\right)^{-\alpha_{\rm h,2}}, r_{\rm h,b}\leq r_{\rm G}<r_{\rm h,t}, \\
	  \left(\frac{r_{\rm h,t}}{r_{\rm h,b}}\right)^{-\alpha_{\rm h,2}}\times\\
	  \left(\frac{r_{\rm G}}{r_{\rm h,t}}\right)^{\epsilon_{\rm h}}\times\\
	  \exp\left(-\frac{r_{\rm G}-r_{\rm h,t}}{\triangle_{\rm h}}\right), r_{\rm G}\geq r_{\rm h,t}, \\
    \end{cases}
    \label{eq:15}
\end{equation}
where $r_{\rm G}$ is the Galactocentric radius in spherical coordinates, and 
$\rho_{\rm H,0}(=5.075\times 10^{-6}\;M_\sun\;\rm pc^{-3})$ is the central mass density of the halo. The cutoff $r_{\rm h,min}=0.5\;\rm kpc$ is to prevent infinite density at Galactic center. 
Other values of parameters in the halo are taken from \citep{kafle2014}, i.e., 
$\alpha_{\rm h,1}=2.4$, 
$\alpha_{\rm h,2}=4.5$, 
$r_{\rm h,b}=17.2\;\rm kpc$, 
$r_{\rm h,t}=97.7\;\rm kpc$, 
$\triangle_{\rm h}=7.1\;\rm kpc$, 
$\epsilon_{\rm h}=\frac{r_{\rm h,t}}{\triangle_{\rm h}}-4.5$. 
The dark matter components have no effect on our results, so only the baryonic mass is considered in the halo.
See \citet{galaxy2016} for further detailed information and discussion. 

The position of the Sun is assumed to be $(R_{\rm sun}, z_{\rm sun}) = (8.5\;\rm kpc,16.5\;pc)$ in the cylindrical coordinates \citep{TheSun1998}. From previous section, we can get the current population of DWDs with ELM WDs for different components of the Galaxy. Then we can get spacial distribution of these binaries according to the spatial density distribution of different components of the Galaxy. 

\section{Results}
\label{sec:4}
\subsection{The number of DDs in the Galaxy}
\label{subsec:4.1}

From the simulations in Section \ref{sec:3}, we can obtain the population of DWDs with ELM WDs in the Galaxy. The present birthrate $\nu$, semi-detached rate $\xi$, total number $N_{\rm tot}$, the contribution of RL channel $\%_{\rm RL}$ and that of the CE channel $\%_{\rm CE}$ are shown in Table \ref{tab:1}. {\color{black}Here ``semi-detached rate" means the rate at which the DWDs become semi-detached. Then these DWDs might evolve into AM CVn systems or directly merge into a single star.} The total number of DWDs with ELM WD companions at 13.7 Gyr is $2.18\times10^7$. The binary systems from RL channel could not be resolved/detected by LISA, due to long orbital periods of such systems ($P_{\rm orb}\gtrsim 0.1\;\rm d$, $f_{\rm GW}\lesssim2\times10^{-4}\;\rm Hz$). However, a part of them would make a contribution to the confusion foreground. For systems from the CE channel, their orbital periods are shorter and more likely to be detected by LISA. 

\startlongtable
\begin{deluxetable}{ccccc}
  \tablecaption{The current birth rate, semi-detached rate, number of DDs in the Galaxy, and the ratio of DDs from different formation channels to the total number \label{tab:1}}
\tablecolumns{6}
\tablenum{1}
\tablewidth{0pt}
\tablehead{
\colhead{} &
\colhead{Bulge} &
\colhead{Disk} & 
\colhead{Halo} & 
\colhead{Galaxy} 
}
\startdata
$\nu\;(10^{-3})$ & $2.74$ & $6.56$ & $0.07$ & $9.37$ \\
$\xi\;(10^{-3})$ & $1.78$ & $4.27$ & $0.05$ & $6.10$ \\
$N_{\rm tot}\;(10^7)$ & $0.63$ & $1.52$ & $0.13$ & $2.18$ \\
$\%_{\rm RL}$ & $65.9\%$ & $65.9\%$ & $82.8\%$ & $67.0\%$ \\
$\%_{\rm CE}$ & $34.1\%$ & $34.1\%$ & $17.2\%$ & $33.0\%$ \\
\enddata
\tablecomments{$\nu = $ current birth rate ($\rm yr^{-1}$), $\xi = $ current semi-detached rate 
($\rm yr^{-1}$), $N_{\rm tot} = $ total number, $\%_{\rm RL}$ and $\%_{\rm CE}$ mean the percentage of the number of systems from RL and CE channels, respectively. 
}
\end{deluxetable}

\begin{figure}
  \centering
  \includegraphics[width=\columnwidth]{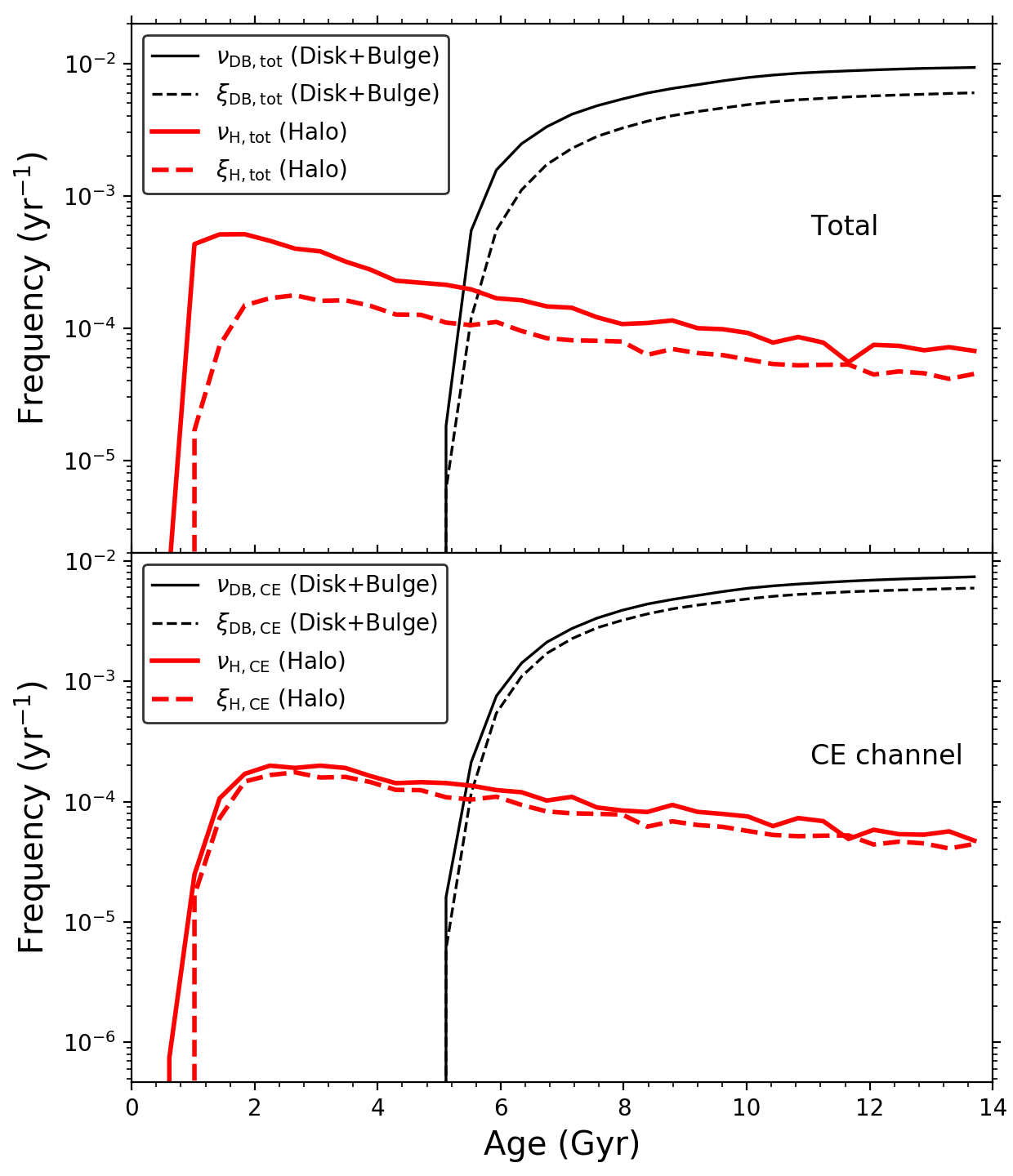}
  \caption{The birthrate and semi-detached rate for the Galaxy components, where the bulge and disk are shown together, due to the similar star formation history (exponential SFR, see Section \ref{subsubsec:3.1.2}). The total birthrate and semi-detached rate are shown in the upper panel, where thin and thick solid lines represent the birthrate for the disk + bulge ($\nu_{\rm DB,tot}$) and the halo ($\nu_{\rm H,tot}$), and the thin and thick dashed lines are for the semi-detached rate of the disk + bulge ($\xi_{\rm DB,tot}$) and the halo ($\xi_{\rm H,tot}$), respectively. In the lower panel, we show the corresponding birthrate and semi-detached rate of DDs with ELM WDs from the CE channel.} 
\label{fig:2}
\end{figure}

As shown in Table \ref{tab:1}, the majority of Galactic DDs with ELM WDs are in the bulge and the disk, where the CE channel contributes more than 30 percent. {However, the contribution of CE channel in the halo is about a half of that in the bulge and disk.} We explain this as follows. In comparison to the bulge and disk, due to low $Z$ of halo stars, the donors (the progenitors of ELM WDs) have large envelope binding energy and short orbital periods (due to small stellar radius) if they fill their Roche lobes at (or near) the base of RGB, making it hard to eject the envelope and form ELM WDs consequently (see also \citetalias{lizw2019}). {In addition, as shown in Figure \ref{fig:1}, ELM WDs in the halo are more likely produced from the RL channel, due to the relatively large parameter space.} Besides, the star formation history also has an effect on this. 

{\color{black}We present the galactic age versus the birthrate and semi-detached rate for different Galaxy components in Figure \ref{fig:2}. The total birthrate and semi-detached rate for the disk + bulge ($\nu_{\rm DB,tot}, \xi_{\rm DB,tot}$) and the halo ($\nu_{\rm DB,tot}, \xi_{\rm DB,tot}$) are shown in the upper panel, while those from the CE channel are presented in the lower panel. 
\begin{figure*}
  \gridline{\fig{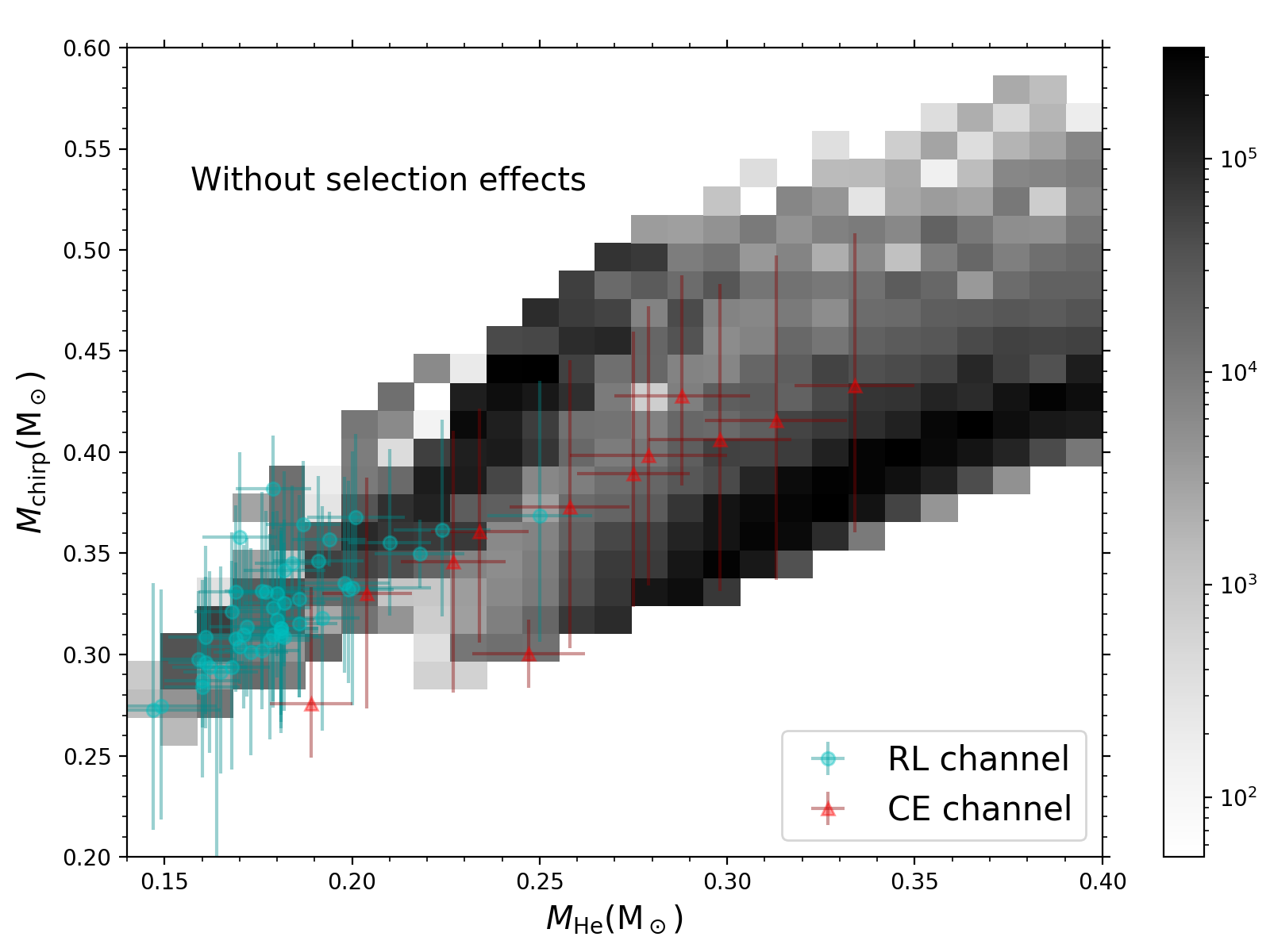}{0.5\textwidth}{}
  \fig{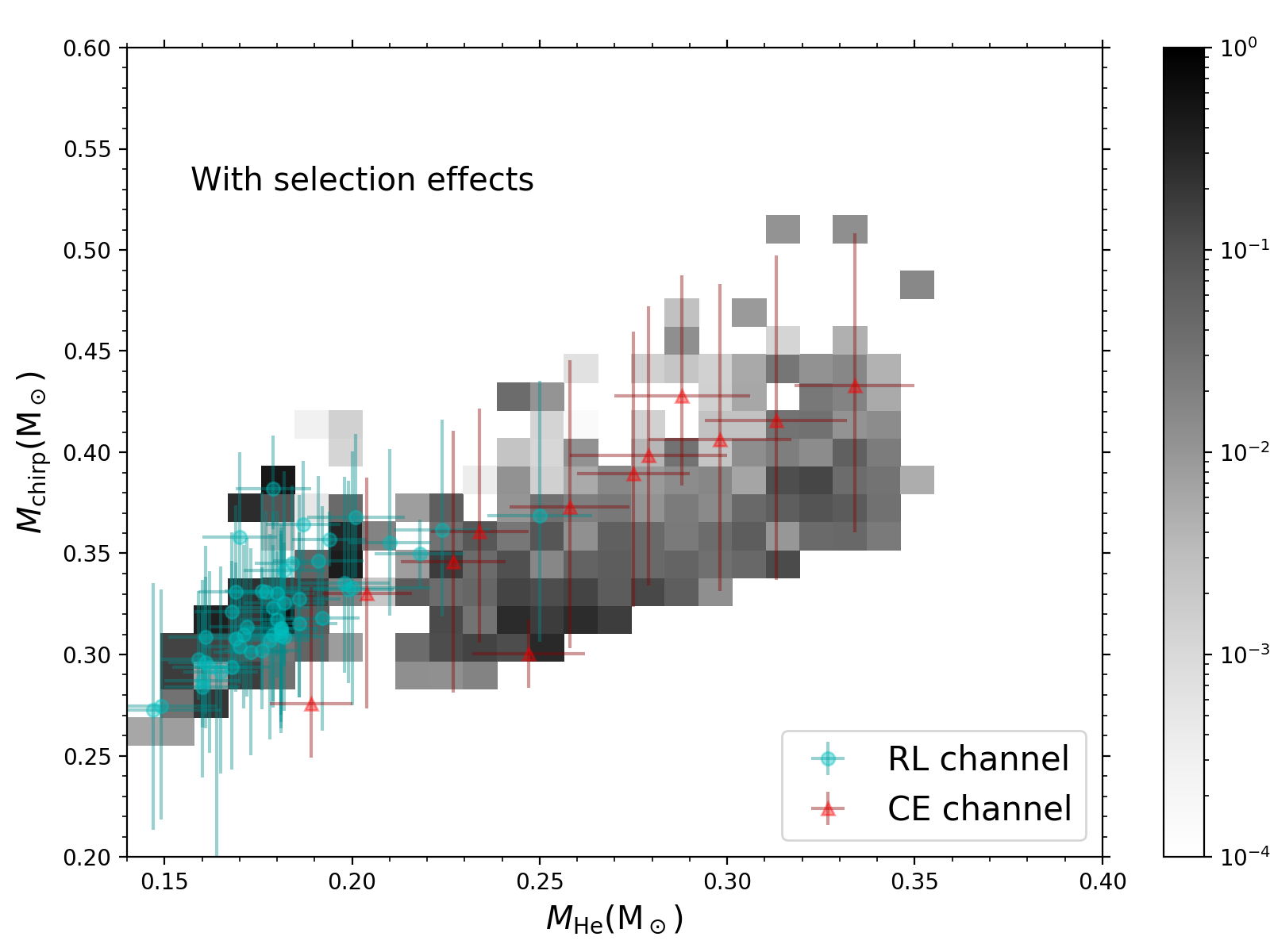}{0.5\textwidth}{}
  }
  \caption{The density distribution in the He WD mass - chirp mass plane for current population of DD with ELM WDs in the Galaxy. The left panel shows all of the DDs with ELM companions without considering the selection effects. The ELM WD mass $M_{\rm He}$ peaks around 0.25 and $0.32M_\odot$ corresponding to systems from RL channel and CE channel, respectively. In the right panel, we take the selection effects in ELM Survey into account, i.e. $k>75\;\rm km\;s^{-1}$, $P_{\rm orb}<2\;\rm d$, $8000<T_{\rm eff}<22000\;\rm K$, $4.85<\log g<7.15$, and magnitude limit. {About 15$\%$ of DDs with ELM WDs ($\sim 3\times10^6$) are retained.} The two peaks of $M_{\rm He}$ change to $0.18$ and $0.25M_\odot$, respectively. {{The observed samples are the clean sample in ELM Survey, and have been divided into two parts according to the formation channel.}} The red triangles are for systems produced from CE channel, and cyan circles are for those formed from RL channel.}
\label{fig:3}
\end{figure*}
In the disk and bulge (exponential SFR), the DWDs with ELM WDs are produced much later than that in the halo, {since the SFR defined in Equation (\ref{eq:8}) indicates that there is no star formation in the first 4 Gyr}. Meanwhile, the birthrate is always larger than the semi-detached rate for the CE channel (see the thin black solid line and dashed lines in the lower panel), indicating that the number of systems from the CE channel in the disk and bulge increases with age. {For the halo, the birthrate and the semi-detached rates for ELM WDs from the CE channel are very close. This is a consequence of the fact that a significant part of systems produced from the CE channel in the halo have become semi-detached or merged due to their extremely short orbital periods.} This result is consistent with the discussion in \citetalias{lizw2019} that most halo objects in ELM Survey are more likely produced from the RL channel (see Figure 11 in \citetalias{lizw2019}). }

\subsection{Chirp mass}
\label{subsec:4.2}

Equation (\ref{eq:2}) shows that the chirp mass determines the evolution of GW frequency of a binary and could be obtained if the variation of GW frequency has been detected. We therefore show in Figure \ref{fig:3} the density distribution of the ELM WD mass versus chirp mass (the $M_{\rm He}-M_{\rm chirp}$ plane) for Galactic DWDs with ELM WDs. {The observed samples are the clean sample in the ELM Survey, which have been divided into two distinct groups according to their formation channels. The red triangles are for those from the CE channel and cyan circles are for those from the RL channel.} {The specific methods to distinguish the formation channel of the observed samples are introduced in \citetalias{lizw2019}. {The basic idea is as follows. We first get a CE efficiency for each sample if we assume that all the observed ELM WDs are from the CE channel. Those with unreasonable values of CE coefficients are considered to be produced from the RL channel, and other samples are supposed to be from the CE channel.}} 

To cover all of the observations, we additionally calculate the parameter spaces for He WDs with mass larger than $0.30\;M_\odot$. The left panel of the figure is for all DDs with ELM WD companions obtained from our calculation, while the right one shows the results with selection effects considered, i.e. $k>75\;\rm km\;s^{-1}$, $P_{\rm orb}<2\;\rm d$, $8000<T_{\rm eff}<22000\;\rm K$, $4.85<\log g<7.15$, and magnitude limit\footnote{The brighter objects are more easier to be detected, then we include the magnitude limit by multiplying a weight of $L_{\rm ELM}^{3/2}$, where $L_{\rm ELM}$ is the luminosity of the ELM WDs.}, as introduced in Section \ref{subsubsec:2.2.1}. 

From Figure \ref{fig:3}, we can find that there are two groups of DWDs. The group with typically lower He WD mass is from RL channel and the other is from CE channel. The typical chirp mass for the RL channel is comparable or even slightly larger than that for the CE channel, since the CO WDs from the RL channel can increase its mass during the mass transfer process. {After the inclusion of selection effects of the ELM Survey, a large part of systems with long orbital periods ($>2\rm d$), and massive He WDs (large $\log g$ and low $T_{\rm eff}$) are removed. The retained systems are about $15\%$ ($\sim3\times10^6$) of the number without considering the selection effects. Besides, the magnitude limit increase the weight of $M_{\rm He}\lesssim0.2M_\odot$.} We find that the low-mass part matches observations well, while the high-mass part is only marginally consistent with observations. This is possibly caused by the assumption of the envelope mass adopted for ELM WDs produced from the CE channel. As we explained in \citetalias{lizw2019}, the envelope mass of ELM WDs from the CE channel is likely less massive than in our study. This means that these WDs should have lower luminosity and larger surface gravity \citep{calcaferro2017}. Therefore, these systems might not be detected by the ELM Survey because of the magnitude limit or upper limit of surface gravity ($\log g =7.15$). Of course, our results may also imply that many He WDs with mass larger than $\sim 0.22\;M_\odot$ are not found yet.

\subsection{The GWR signals of DDs with ELM WDs}
\label{subsec:4.3}

\begin{figure}
    \centering
    \includegraphics[width=\columnwidth]{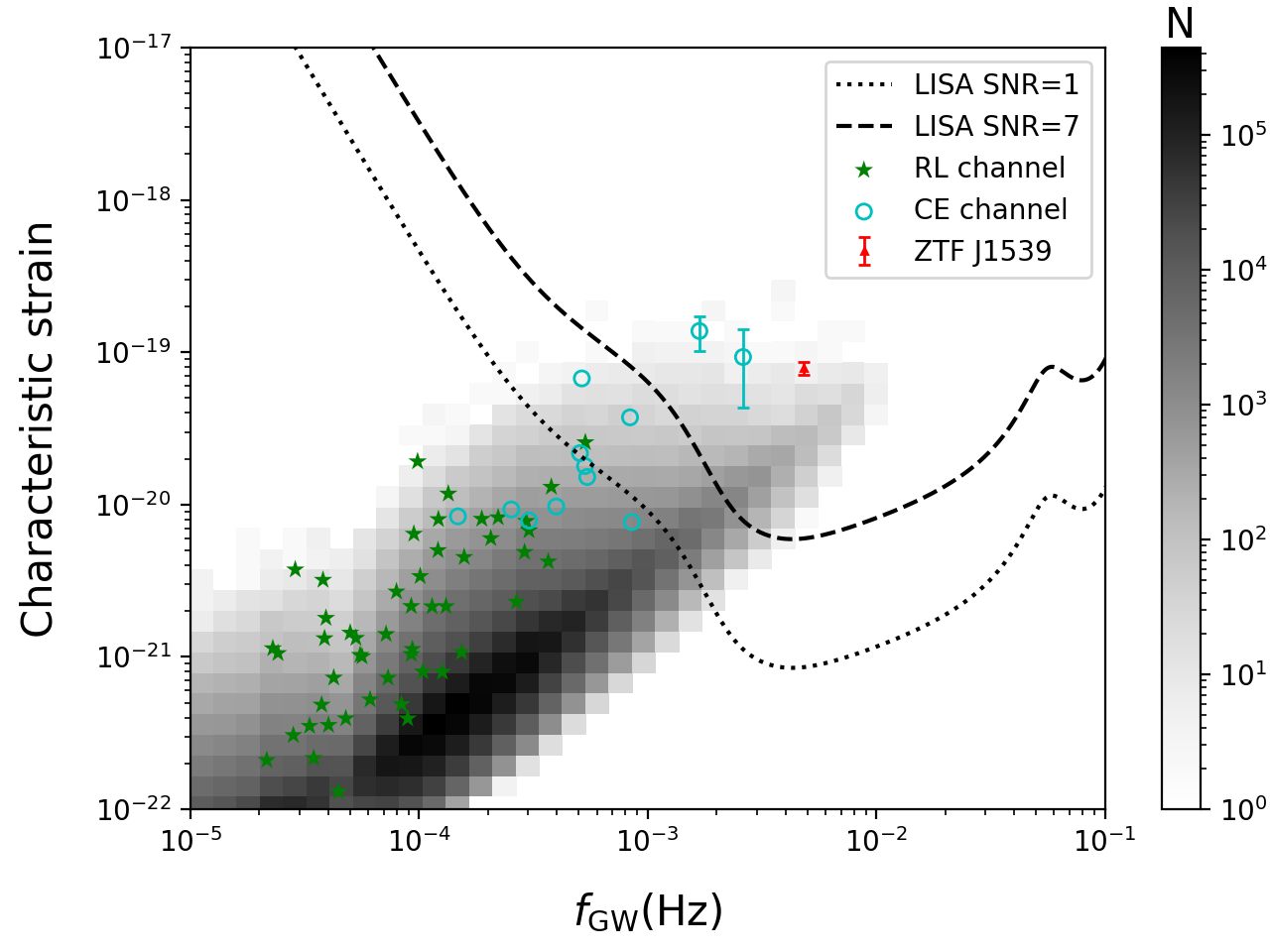}
    \caption{The GW characteristic strains of current population of DDs with ELM WD companions in the Galaxy. The observed samples from ELM Survey are shown with green filled stars and cyan open circles corresponding to systems from RL channel and CE channel, respectively. The dotted and dashed lines are the sensitivity curves with SNR = 1 and 7, respectively. The error bars of J0651 and J0935 are taken from \citet{kupfer2018}, and ZTF J1539 is an eclipsing DWD system with an orbital period of 6.91 minutes, its GW strain is taken from \citet{nature2019}. 
    }
    \label{fig:4}
\end{figure}

Figure \ref{fig:4} shows the GW frequency versus the characteristic strain of DDs with ELM WDs, where the dotted and dashed lines are the all-sky averaged sensitivity curves of \texttt{LISA} when SNR = 1 and 7, respectively. {The characteristic strains of observed samples are also presented in the figure for comparison, and the three with error bars are the potential verification sources for \texttt{LISA} \citep{kupfer2018,nature2019}}. It is shown that the two distinct populations separated by ELM WD masses in Figure \ref{fig:3} are mixing together and become undistinguishable due to their similar chirp masses and continuous orbital period distribution as shown in the bottom panel of Figure 13 in \citetalias{lizw2019}. For systems from RL channel, they generally have large orbital periods, i.e. $f_{\rm GW}\lesssim 10^{-4}\;\rm Hz$. On the contrary, the high frequency region is dominated by DDs from CE channel.

{It appears that the characteristic strain $h_{\rm c}$ from the observed samples is consistent with the theoretical prediction, but is larger than the characteristic strain in the most dense region. We verify that the most dense region in Figure \ref{fig:4} arises from ELM WDs produced via the CE channel.} Due to the uncertainty of the CE process, we assume the ELM WDs have the same structures as the ones from the RL channel and find that many ELM WDs in DDs with $M_{\rm ELM}\gtrsim 0.22M_\odot$ are from the CE channel in \citetalias{lizw2019}. These relatively massive ELM WDs have not been confirmed yet \citep{pelisoli2019b}, but they are the dominate sources in Figure \ref{fig:4}. See also the discussions in Section \ref{subsec:4.2}. 

{We confirm that the three most compact systems \textsf{J0651}, \textsf{J0935} and ZTF J1539 with orbital period about 765 s, 1188 s, and 414 s, should be detected by LISA with SNR 94, 48 and 130, respectively. These values are close to those in \citet{kupfer2018,nature2019}, with SNR of 90, 45, 143, respectively. The minor differences may result from the different sensitivity curves (confusion noise and instrument noise) we adopt.} The detailed discussion of the combination of EM and GW observations for the three objects will be presented in Section \ref{subsec:4.5}. For systems with gravitational wave frequency below $\sim 1$ mHz, the GW signals of these binaries will make up the confusion foreground noise (see Section \ref{subsec:4.4}). 

\subsection{Foreground noise and resolved sources}
\label{subsec:4.4}

\begin{figure}
    \centering
    \includegraphics[width=\columnwidth]{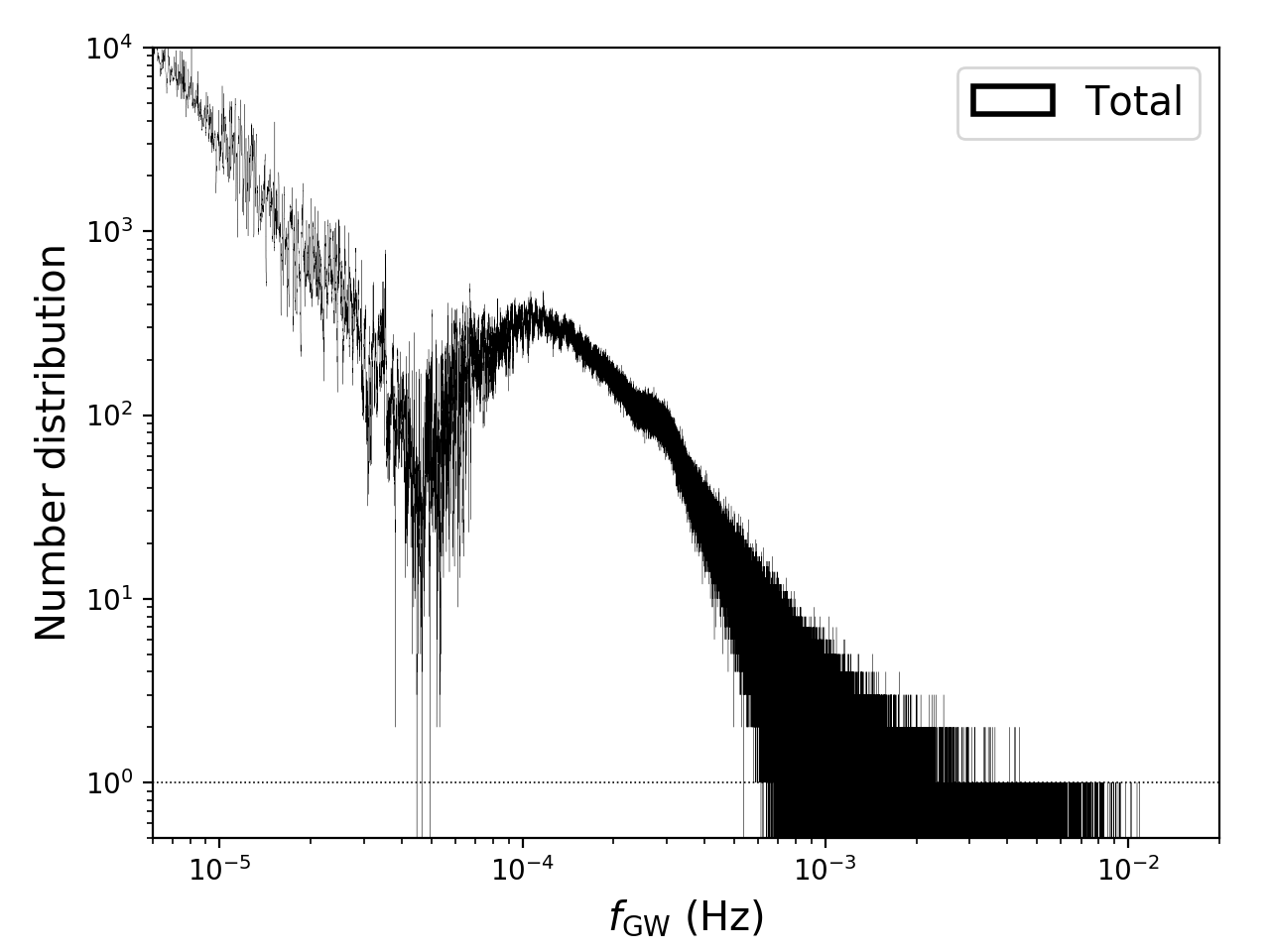}
    \includegraphics[width=\columnwidth]{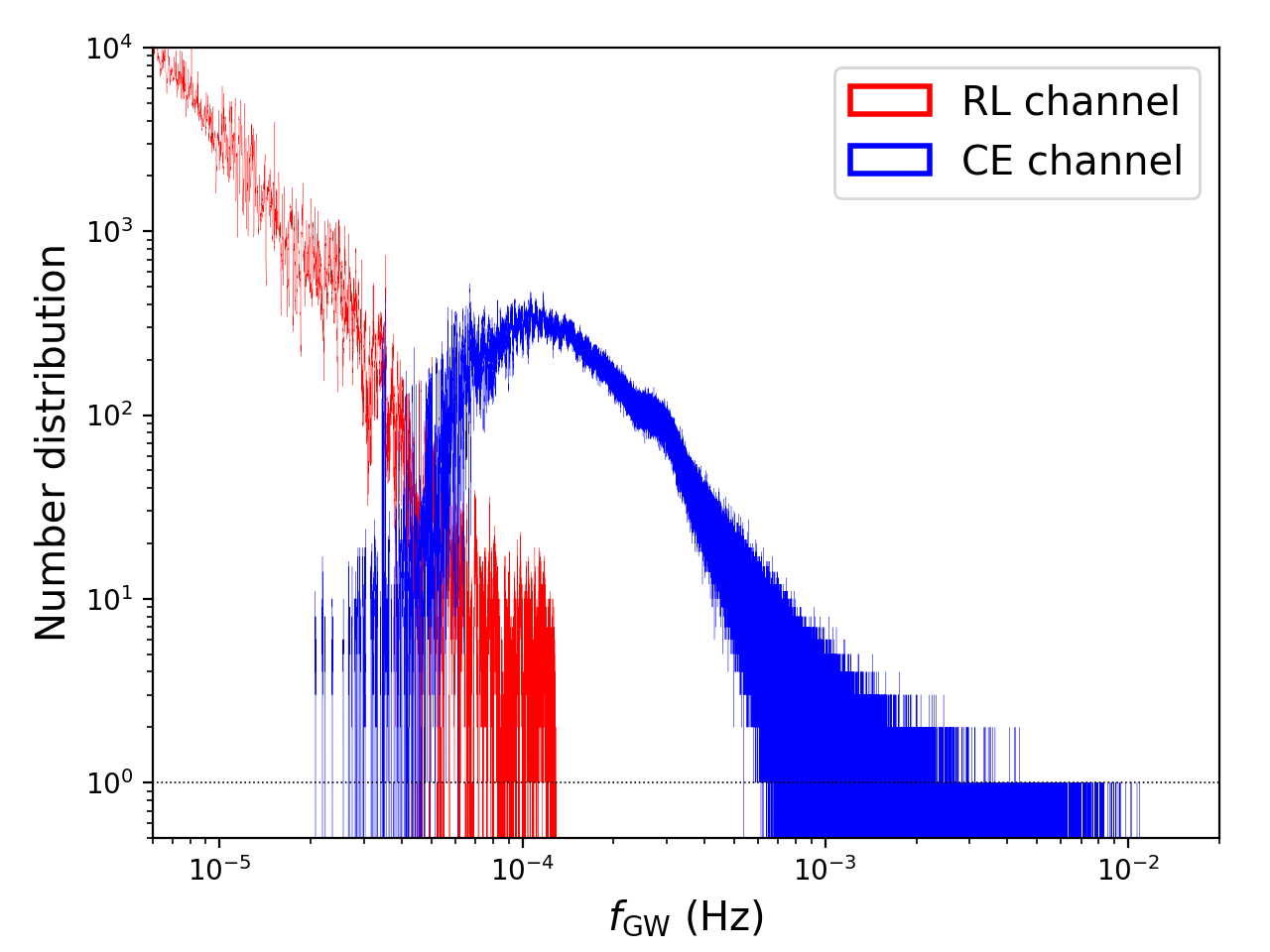}
    \caption{{The upper panel shows the number distribution of all ELM DDs in the Galaxy per frequency bin ($\triangle f_{\rm GW} = 1/4 \;\rm yr^{-1}$), and the contributions from different formation channels are shown in the lower panel.} The sources could be resolved for frequency larger than $\sim 0.6\;\rm mHz $. The distribution shows two distinct groups with ELM WDs from the RL channel ($f_{\rm GW}\lesssim 10^{-4}\;\rm Hz$) and CE channel ($f_{GW}\gtrsim 4\times 10^{-5}\;\rm Hz$), respectively. The dispersion of number distribution around $5\times 10^{-5}\;\rm Hz$ is caused by the small number of ELM WDs from both channels here. See texts for more details. }
    \label{fig:5}
\end{figure}

\begin{figure*}
    \centering
    \includegraphics[width=0.8\textwidth]{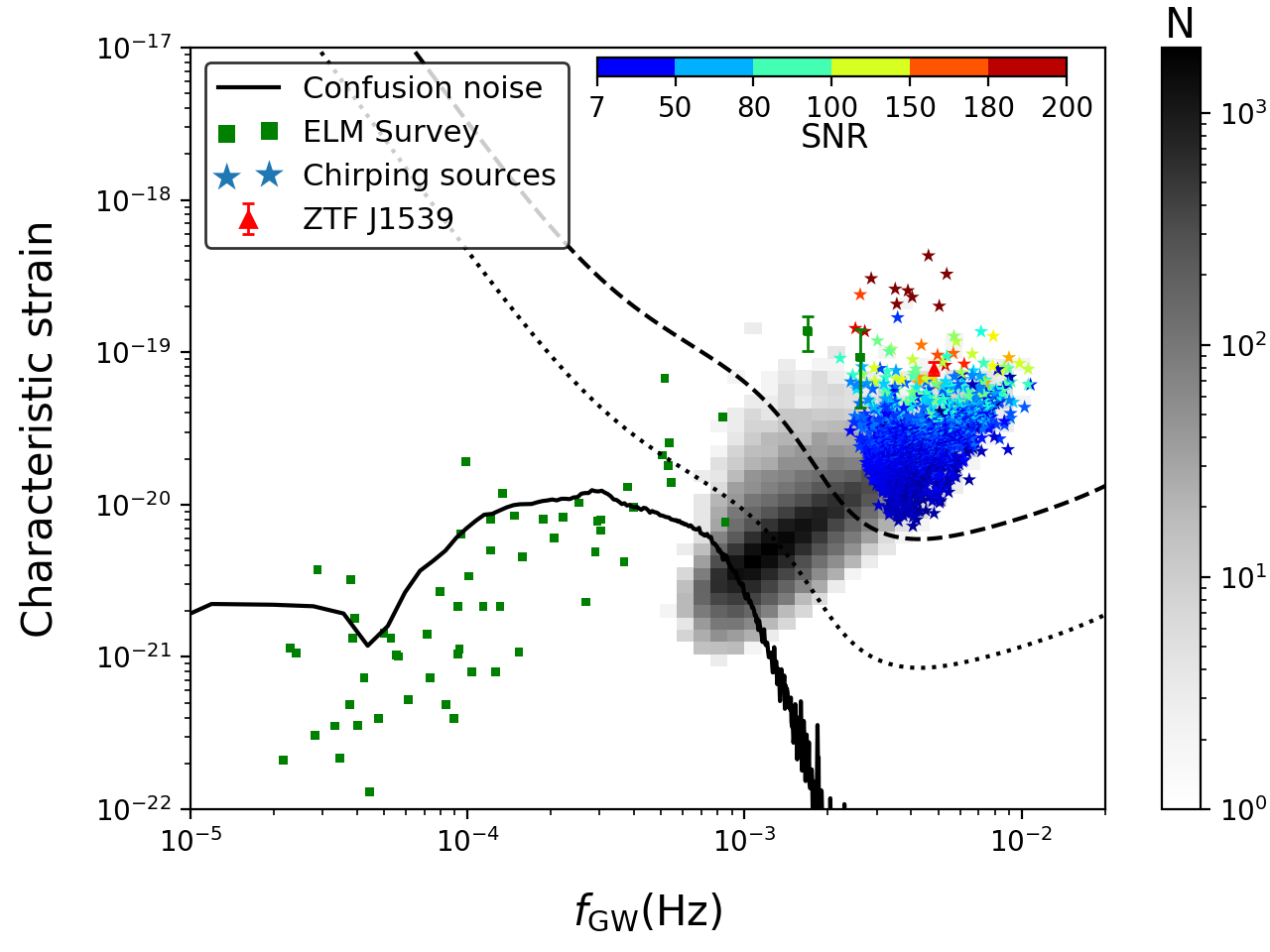}
    \caption{The foreground noise and resolved sources of ELM DDs. The LISA sensitivity curves of SNR = 1, 7 are shown in dotted and dashed lines, respectively. The samples in ELM Survey are shown in green squares. The resolved sources are shown in grey scale, among which 6243 sources are above the LISA sensitivity curve with SNR $= 7$. The color stars represent the 2023 chirping sources and the colors label the SNR of these systems. The confusion foreground noise is shown in solid line with a running average of 1000 bins. 
    }
    \label{fig:6}
\end{figure*}

To study the GW foreground noise and the resolved sources from DWDs with ELM WDs, we show the GW frequency versus the number distribution of such objects in the upper panel of Figure \ref{fig:5}, where the width of each bin is chosen as $\triangle f_{\rm GW} = 1/T_{\rm obs} = 7.93\times10^{-9}\;\rm Hz$. Sources are called resolved sources if only one individual source in one frequency bin. Here the ``resolved sources" only involves the LISA integration time ($T_{\rm obs} = 4 \;\rm yr$), and whether these sources could be detected by LISA will be discussed below. 

We see that the first resolved source appears at $f_{\rm GW} \sim 6\times10^{-4}\;\rm Hz$, which is smaller than the value ($1.5-2.5\;\rm mHz$) given by \citet{nelemans2001a} (see also \citealt{ruiter2010}), because (1) we adopt a longer mission time of LISA i.e. $t_{\rm obs}=4\;\rm yr$ instead of 1 yr; (2) we only focus on the DWDs with ELM WDs in this study while \citet{nelemans2001a} were mainly interested in DWDs with relatively massive WDs. We should bear in mind that not all sources with $f_{\rm GW}$ greater than the critical value could be resolved. It can be seen in Figure \ref{fig:5}, that there are more than one sources in one bin even when $f_{\rm GW}\gtrsim 10^{-3}\;\rm Hz$. In this case, the very bright sources in these bins could be extracted from the LISA data \citep{littenberg2011}.

{The number distribution shows two distinct groups with ELM WDs from the RL channel ($f_{\rm GW}\lesssim 10^{-4}\;\rm Hz$) and CE channel ($f_{\rm GW}\gtrsim 4\times 10^{-5}\;\rm Hz$), respectively, as shown in the lower panel of Figure \ref{fig:5}.} The dispersion is very large at $5\times 10^{-5}\;\rm Hz$ ($P_{\rm orb}\sim 0.5\;\rm d$) in the total number distribution. This is because the numbers of ELM WDs from both of the RL channel and CE channel at this frequency are very small. In RL channel, the parameter space to produce systems with orbital period less than $\sim 0.5\;\rm d$ is very small. Few systems have $f_{\rm GW}$ larger than $5\times 10^{-5}\;\rm Hz$. In CE channel, the small number at $f_{\rm GW}\lesssim 5\;\rm mHz$ is because that only ELM WDs with mass larger than $\sim 0.27M_\odot$ can eject the CE successfully producing systems with final periods $\gtrsim 0.5\;\rm d$. If we consider the contribution of other types of DWDs, e.g. CO WD + CO WD, the dispersion shown in Figure~\ref{fig:5} may be removed (as shown by \citealt{yus2010}). A comprehensive study of GWR from the whole DWD population is necessary and will be given in the next paper where we will include recent progresses in the study of dynamically instability and the CE evolution in binary evolution as well.  

In each frequency bin, we calculate the net GW amplitude as \citet{timpano2006}
\begin{equation}
  h_{\rm net} = \left(\sum_{i=1}^{N_{\rm b}}(h_{\rm c}^2)_i\right)^{1/2},
  \label{eq:16}
\end{equation}
where $N_{\rm b}$ is the number of sources in each bin. With average of 1000 bins of net amplitude we obtain a smoothing foreground noise shown in Figure \ref{fig:6}. In the plot, the distribution of resolved sources is also shown by the grey scale. Both the number of GW sources and individual GW strains determine the shape of the foreground noise. With the increase of $f_{\rm GW}$, the foreground noise remains basically unchanged at $f_{\rm GW}\lesssim 4\times10^{-5}\;\rm Hz$ due to the increase of the GW strains of individual sources and the decrease of number of sources (see Figure~\ref{fig:5}). The dip at $f_{\rm GW}\approx4.4\times 10^{-5}\;\rm Hz$ is caused by the small number of ELM WDs from both channels here, as explained above. Then the foreground noise increases from $\sim 5\times 10^{-5}\;\rm Hz$ to $3\times 10^{-4}\;\rm Hz$ due to the increase of the GW strains of individual sources, but falls down when $f_{\rm GW} \gtrsim 3\times10^{-4}\;\rm Hz$ because of sharply decreasing of the number of GW sources.


\begin{figure*}
	\centering
    \includegraphics[width=0.8\textwidth]{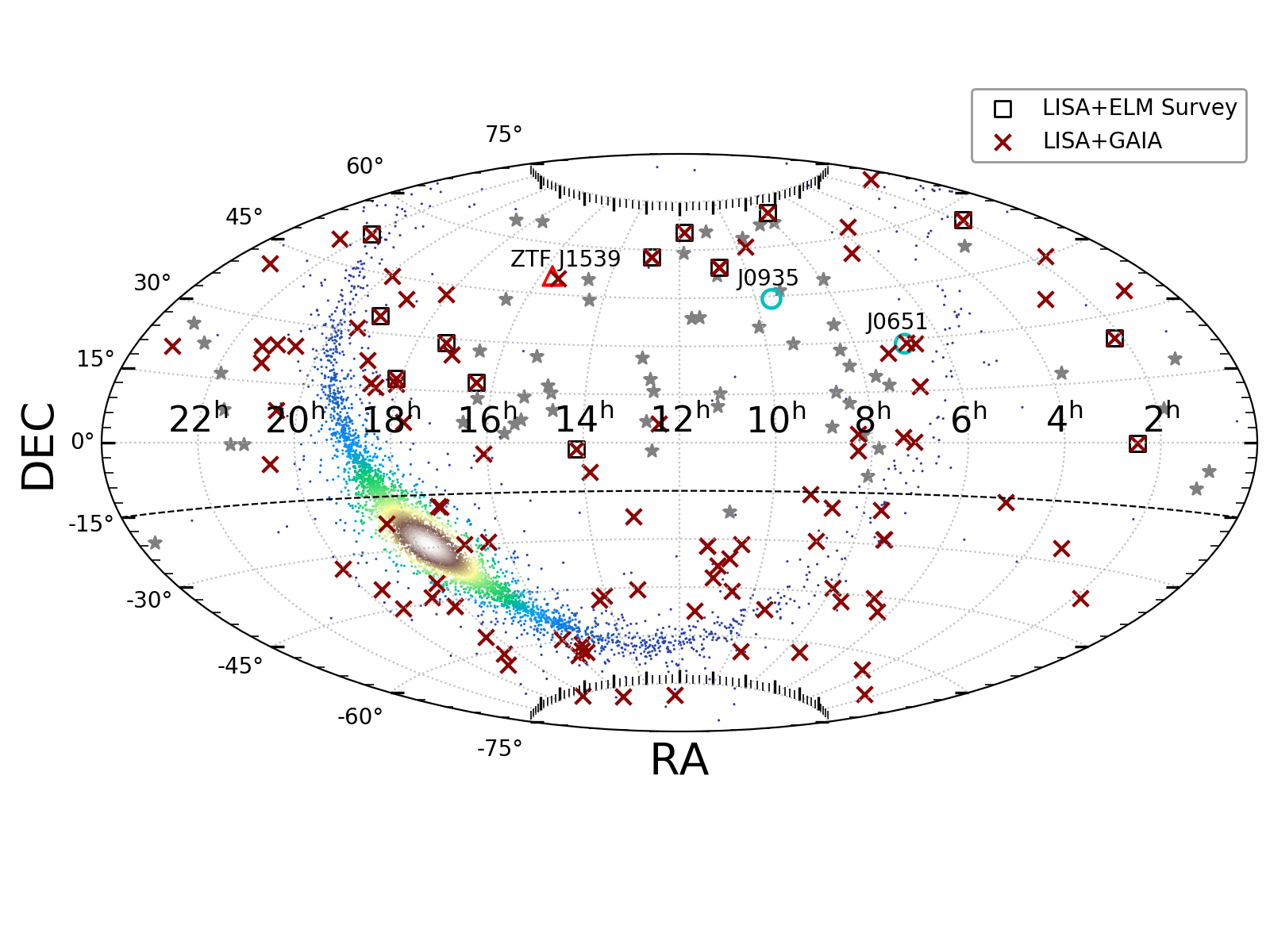}
    \caption{The sky positions of LISA detectable sources (color dots) and the clean sample of ELM Survey (grey filled stars). These sources with black squares also satisfy the selection effects in ELM Survey. These sources with red crosses are expected to be observed by the {\it{Gaia}} space mission. The two cyan open circles and the red open triangle are the verification sources for LISA. 
    }
    \label{fig:7}
\end{figure*}

We totally obtain $45735$ resolved sources, among which 6243 sources\footnote{{The number is slightly larger than that of the recent work of \citet{lamberts2019} on GWR of DWDs, who found 5600 resolved sources of CO + He WD systems with SNR $>7$. The explanation for this difference is due to the different initial binary fraction we used. In this work, the systems with orbital periods less than 100 yrs are assumed to be binaries. However, \citet{lamberts2019} assumed that the orbital separation distribution is uniform in $\log a$ for a $\leq10^6R_\odot$, and a binary fraction of 50 percent was additionally adopted. Consequently, the initial binary fraction in our work is larger than that assumed in \citet{lamberts2019}, which leading to the large number of LISA sources in our results.}} are above the LISA sensitivity curve with SNR = 7.  Some resolved sources have very large SNR and could be detected with chirp signals, i.e., the time derivative of GW frequency $\dot{f}_{\rm GW}$ is larger than the minimum observable chirp $\dot{f}_{\rm GW,min}$, which is written as \citep{takahashi2002,seto2002}
\begin{equation}
  \begin{split}
	\dot{f}_{\rm GW,min} &\sim C\times \triangle \dot{f}_{\rm GW}\\
	&\sim C\times \frac{6\sqrt{5}}{\pi} \frac 1 {T_{\rm obs}^2}\frac 1 {\rm SNR}\\ 
	&\approx  1.3\times10^{-17}\left(\frac{100}{\rm SNR}\right)\
	\left(\frac{4 \;\rm yr}{T_{\rm obs}}\right)^2 \rm Hz\;s^{-1},
  \end{split}
  \label{eq:17}
\end{equation}
where $C$ was assumed to be 5, meaning that the chirp signal is with 20 percent accuracy in the 
LISA measurement (see also \citealt{tauris2018}). 
The color stars in Figure \ref{fig:6} represent the 2023 chirping sources obtained in our simulation. Different colors in the plot indicate different SNR values. The chirping signals only appear at relatively high GW frequencies (i.e. $f_{\rm GW}\gtrsim 2\times 10^{-3}\;\rm Hz$), mainly due to the high sensitivity of LISA designed in this frequency range and relatively strong GW strains of individual sources with these frequencies (see Figure \ref{fig:6}).

\subsection{Combination of EM and GW observations}
\label{subsec:4.5}

We have already shown in Section \ref{subsec:4.3} that the three most compact DWDs with ELM WDs would be very likely verified by LISA in the future. This indicates that a few DWDs with ELM WDs could be detected by GW detection and EM observations as well. The combination of GW and EM observations could improve the accuracy of distance measurement and constrain the Galaxy structure further \citep{shah2012,shah2013,korol2019,littenberg2019}. In this part, we give some discussions on EM detections for the detected GW sources of DWDs with ELM WDs. 

{We consider two projects for EM observations, i.e. the ELM Survey and the {\it{Gaia}} observation. The ELM Survey gives well defined selection effects of ELM WDs, and can be used to test our simulation results. {\it{Gaia}} has the potential to discover an unbiased sample of \texttt{LISA} verification sources \citep{korol2017}, and will give a more complete sample of ELM WDs. In order to get the detectable samples of these two projects, we firstly assume the ELM WDs emit the blackbody spectrum. Then redden the flux using the \citet{fitzpatrick1999} parameterization, where the integrated extinction in the direction of the sources is taken from \citet{SFD1998}\footnote{{These calculations are done with the \texttt{Python} package of \texttt{dustmaps}} (\url{http://argonaut.skymaps.info}, \citealt{green2018}). }. Finally, we obtain the visual magnitude of the ELM WDs in SDSS $g-$band ({\it{Gaia}}$ \;G-$band for {\it{Gaia}}) for each resolved GW sources by using the $g-$band filter of SDSS ($G-$band filter of {\it{Gaia}}), as well as the distances obtained by the Galaxy model.} For ELM Survey, we also consider the selection effects introduced in Section \ref{subsubsec:2.2.1}. {The limiting magnitude is set to be 20 for ELM Survey \citep{brown2010}, and 21 for {\it{Gaia}} \citep{GAIA2016}}. Since 6.5m MMT is only focusing on the northern sky, we assume that the sources with DEC $>-15^\circ$ could be detected by the ELM Survey for simplicity. {There is no sky position limit for {\it{Gaia}}.} 

Figure \ref{fig:7} shows the sky positions in an equatorial coordinate for LISA detectable sources. The black squares and red crosses represent these sources that expected to be detected by ELM Survey and {\it{Gaia}}, respectively. The 62 DWDs with ELM WDs obtained by the ELM Survey are presented with grey filled stars for comparison. Almost all GW detectable sources are in the disk and bulge as expected. Since the objects in the halo are much less than that in the disk and bulge, and they are mainly from the RL channel (relatively long orbital period, see Table \ref{tab:1}). Therefore, the GW signals from the halo have a negligible effect for LISA detection (see also \citealt{ruiter2009}). {The expected EM counterparts for those potential GW sources are all in the disk. Although about $1/3$ observed samples in ELM Survey are in the halo, these sources are not likely to be detected by LISA. Because most of them are more likely produced from the RL channel due to the relatively longer orbital period ($P_{\rm orb}\gtrsim0.1\;\rm d$) and smaller ELM WD masses ($M_{\rm ELM}\lesssim 0.2M_\odot$).} The expected numbers under different combinations of EM and GW observations are summarized in Table \ref{tab:2}. 

There are 13 binaries satisfying the selection effects for the ELM Survey, after we limit the observed region of ${\rm DEC>-15^\circ}$. Besides, we found that 107 sources have the possibility to be observed by {\it{Gaia}}. And more than a half (59) of them are in the region of $\rm DEC<0^\circ$. This result is consistent with the discussions in \citet{kupfer2018} that most LISA verification binaries in the Northern hemisphere are caused by the incomplete and biased samples in the observations, and more unbiased sources at low-Galactic latitudes are expected to be detected in the future. 

\startlongtable
\begin{deluxetable}{lc}
  \tablecaption{The statistical results of LISA sources \label{tab:2}}
\tablecolumns{6}
\tablenum{2}
\tablewidth{0pt}
\tablehead{
\colhead{Constraints} &
\colhead{Number}  
}
\startdata
LISA  & 6243\\
LISA + Chirp & 2023\\
LISA + ELM Survey & 13\\
LISA + {\it{Gaia}} & 107\\
LISA + {\it{Gaia}} + Chirp & 18\\
\enddata
\tablecomments{``LISA" means sources with $\rm SNR > 7$, ``Chirp" means that the sources have measurable chirping signals, and its definition could be found in Section \ref{subsec:4.4}.
}
\end{deluxetable}

\begin{figure}
    \centering
    \includegraphics[width=\columnwidth]{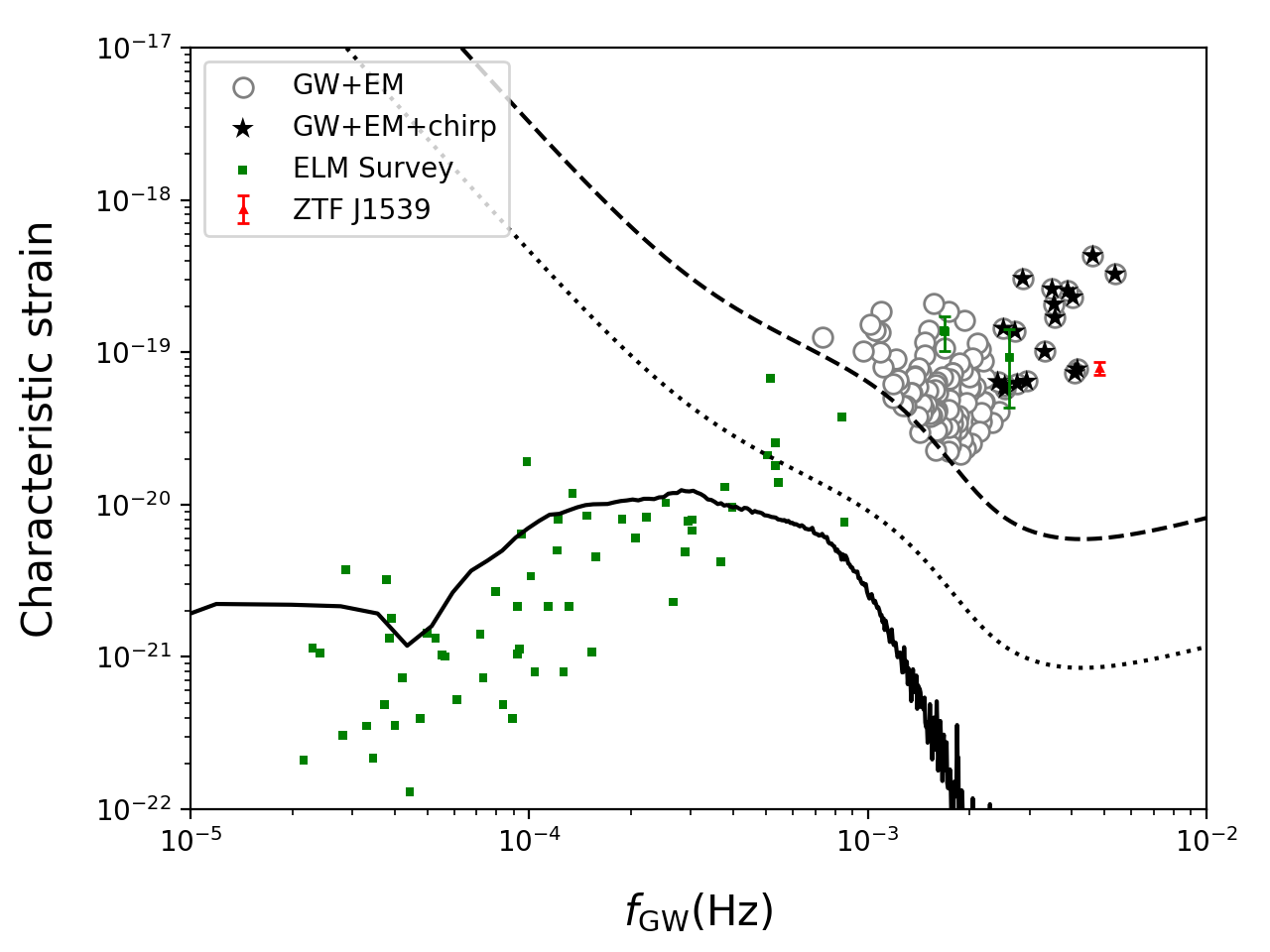}
    \caption{The distribution of possible ELM WDs from the combination of EM ({\it{Gaia}}) and GW observations in $f_{\rm GW}-$characteristic strain plane. The observed samples in ELM Survey are shown in green filled squares. The 107 detectable and 18 chirping sources are shown in grey circles and black stars, respectively. 
    }
    \label{fig:8}
\end{figure}

We present the characteristic strain of possible candidates from the combination of LISA and {\it{Gaia}} observations in Figure \ref{fig:8}, where the green filled squares are the clean samples in ELM Survey, the black open circles are sources in our simulations that could be detected by both GW and EM observations, and the black filled stars are for the chirping sources. There are totally $18$ sources with measurable chirping signals. According to Equation (\ref{eq:17}), J0651 and ZTF J1539 have observable chirping signals due to their short orbital periods, but it is hard to detect such signal for J0935. 

\begin{figure}
    \centering
    \includegraphics[width=\columnwidth]{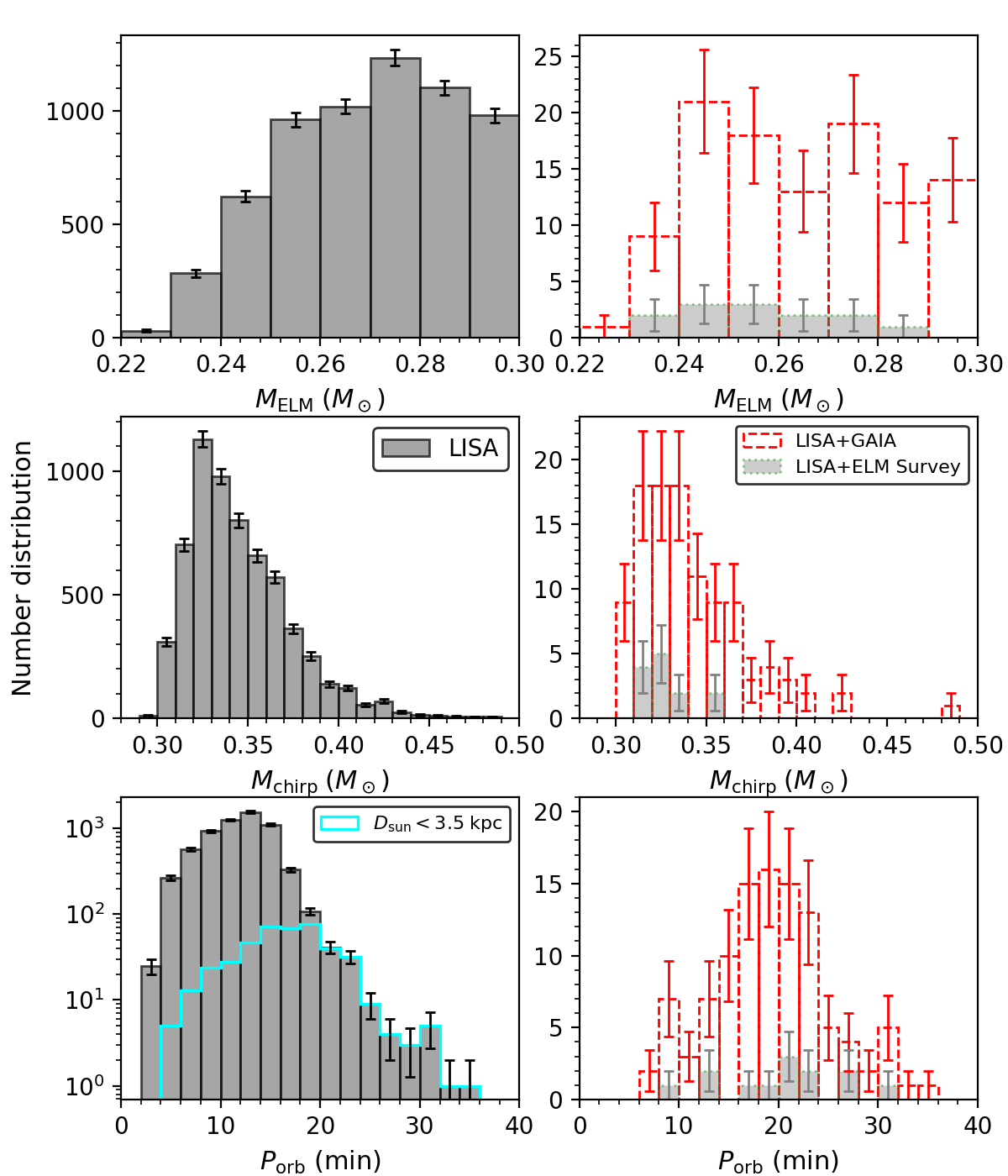}
    \caption{ELM WD mass, chirp mass and orbital period distribution of possibly resolved DWD candidates for LISA observations and combination of EM and GW observations. The LISA sources with distance, $D_{\rm sun}$, less than $3.5\;\rm kpc$ are shown in cyan hatched region. The error represents the Poisson error. 
    }
    \label{fig:9}
\end{figure}

Figure \ref{fig:9} shows distributions of ELM WD mass, chirp mass and orbital period for the sources which could be detected by LISA (SNR$>7$, the left panels) and the detected sources from the combined observation of LISA and {\it{Gaia}}/ELM Survey (the right panels). From Figure \ref{fig:9}, we can find that the ELM WD mass distribution peaks around $\sim0.27M_\odot$ and the chirp mass distribution peaks around $0.33\;M_\odot$. These peaks are located at smaller mass compared with Figure \ref{fig:3}. This can be understood as follow. As shown in Section \ref{subsec:4.3} and \ref{subsec:4.4}, all the LISA resolved sources are produced from the CE channel. These sourcecs with larger ELM WD mass and chirp mass have larger orbital periods after the CE process. Then the GW frequency of these sources is too low to be resolved by LISA. 

{Most of LISA detectable sources have short orbital periods ($\lesssim 20$ min) as shown in the lower left panel. This can be understood as follows. First, the amplitude of GW signals increase with the frequency. Therefore these sources with short orbital periods are more likely to be detected with LISA. Secondly, the LISA is more sensitive to frequency at a few mHZ. Given that the amplitude of GW signals scales as $1/d$ and the amplitude of EM signals scale as $1/d^2$, GW detectors can detect much further than EM detectors. To see the difference between the EM and GW observations, we make a cutoff at $3.5\;\rm kpc$ (see lower left panel), which is considered as the maximum observed distance of {\it{Gaia}} \citep{korol2017}. From this plot, we can find that these sources with shorter orbital periods can be detected by LISA but not EM observation. This explains why these detectable sources of the combination of GW and EM observations have a larger typical orbital period around 20 min compared with detectable sources from GW observations.}

\section{Discussions}
\label{sec:5}
\subsection{The uncertainties in the CE process}
\label{subsec:5.1}
One of the main uncertainties of our results comes from the envelope structure of ELM WDs produced from the CE channel, which is still an open question for the CE ejection process. And then we assumed that the envelope structure of these ELM WDs is similar to that of ELM WDs with the same mass from RL channel. If we overestimate the envelope thickness of ELM WDs from the CE channel, then some systems would not be detected by optical telescope, due to the large surface gravity and low luminosity \citep{calcaferro2017}. 
{The combination of the EM and GW observations are expect to give a constraint of the envelope structure of ELM WDs from the CE channel. For example, the GW strain contains the information of distance to the sources, which can further help to verify the estimates of radius and surface temperature in the spectroscopic analysis \citep{burdge2019}. These physical quantities are important to infer the lifetime of ELM WDs, which has a strong correlation to the envelope structure of ELM WDs \citep{althaus2013,istrate2014b}. It is worth noting that all LISA resolved sources undergo at least one CE phase \citep{yus2010,ruiter2010}. Therefore, the statistical properties of GW sources can also limit the CE parameters.}

The assumption of CE ejection process have a large effect on our final results. In this work, we adopt the standard energy prescription to simplify this process, with $\alpha_{\rm CE}\lambda=1$. {{This parameter value can reproduce the mass distribution of ELM WDs with mass less than $\sim0.2M_\odot$ as well as the companion star mass distribution.}}
Here we test the effect of different parameters in the CE phase, e.g. $\alpha_{\rm CE}\lambda=0.25,0.5$. The corresponding results are shown in table \ref{tab:3}.  The standard model with $\alpha_{\rm CE}\lambda=1$ is shown for comparison. We can see that the proportions of systems from CE channel are strongly correlated with the CE coefficient, since CE can be ejected more efficiently with larger $\alpha_{\rm CE}\lambda$ (See also \citetalias{lizw2019}). For the cases of $\alpha_{\rm CE}\lambda=0.25,0.5$, most of the ELM WDs are produced from the RL channel, which has less contribution to the LISA detection. We find about $382$ and $987$ sources have SNR larger than 7, and only 7 and 18 systems could be detected by the combination of EM and GW observations for these two cases, respectively. 

\startlongtable
\begin{deluxetable}{lccccc}
  \tablecaption{The effects of CE parameters\label{tab:3}}
\tablecolumns{6}
\tablenum{3}
\tablewidth{0pt}
\tablehead{
\colhead{$\alpha_{\rm CE}\lambda$} &
\colhead{$N_{\rm tot}$} &
\colhead{$\%_{\rm CE}$} & 
\colhead{LISA} &
\colhead{Chirp} &
\colhead{LISA+{\it{Gaia}}}
}
\startdata
$0.25$ & $1.48\times10^7$ &  $0.2\%$ & $382$ & $130$ & $7$\\
$0.5$ & $1.47\times10^7$ &  $5.0\%$ & $987$ & $303$ & $18$\\
$1$ & $2.18\times10^7$ &  $34.1\%$ & $6243$ & $2023$ & $107$\\
\enddata
\tablecomments{The signs are same as labeled in table \ref{tab:1} and table \ref{tab:2}.
}
\end{deluxetable}

\subsection{The influence of Galaxy model on the results}
\label{subsec:5.2}
To test the influence of different Galaxy models on the results, we use similar methods to calculate the GWR of DDs with ELM WD companions in the Galaxy model firstly suggested by \citet{bp1999} (has been developed by \citealt{nelemans2004}, BP99 model hereafter). For convenience, we only consider the disk and the bulge components in this model, since the contribution of GWR from the halo populations is expected to be small, as discussed above (see also \citealt{ruiter2009}). The total mass of the Galaxy is assumed to be $6.8\times10^{10}\;M_\odot$, with bulge mass and disk mass of $2\times10^{10}\;M_\odot$ and $4.8\times10^{10}\;M_\odot$ for the sake of comparison with the Galaxy model (Fiducial model) assumed in Section \ref{sec:3}. The SFR is taken from \citet{bp1999}, with doubling the SFR in the inner 3 kpc for the bulge (\citealt{nelemans2004}, see also \citealt{korol2017}). This model gives the Galactic age of 13.7 Gyr, with continuous star formation in the Galaxy history. However, in the fiducial model, {the age of disk and bulge is assumed to be 9.7 Gyr, as shown in Equation (\ref{eq:8})}. 

With BP99 model, we can get $2.92\times10^{7}$ DDs with ELM WD companions, 4408 LISA sources, 1305 chirping sources and 72 candidates for LISA+{\it{Gaia}} observations (see table \ref{tab:4}). Comparing with our Fiducial model, we find that the total number of DDs with ELM WD companions in BP99 model is larger than that in the fiducial model. This difference mainly comes from different star formation histories in these two models (see also \citealt{yus2013}). In BP99 model, since star formation is continuous over 13.7 Gyr, a considerable part of ELM WDs are produced from progenitors with mass less than $1.1\;M_\odot$ (with main sequence lifetime larger than $\sim 9\;\rm Gyr$), which are numerous based on the IMF. However, in the fiducial model, due to {the shorter age of disk and bulge (9.7 Gyr)}, most of ELM WDs are produced from more massive progenitors. Besides, for ELM WDs from the CE channel, after the ejection of CE, the final orbital periods of ELM WDs with low-mass progenitors ($\lesssim 1.1\;M_\odot$) are longer compared with that of the massive progenitors, which requires more orbital energy from the orbit shrinking to eject the thick envelope. Therefore, In the BP99 model, relatively less number of systems could be detected by LISA, as well as the combination with the EM observation. 

\startlongtable
\begin{deluxetable}{ccccc}
  \tablecaption{Comparison with different Galaxy model\label{tab:4}}
\tablecolumns{4}
\tablenum{4}
\tablewidth{0pt}
\tablehead{
\colhead{Galaxy Model} &
\colhead{$N_{\rm tot}$} &
\colhead{LISA} & 
\colhead{Chirp} & 
\colhead{LISA+{\it{Gaia}}}  
}
\startdata
BP99 model& $2.92\times10^7$ &  $4048$ & $1305$ & $72$ \\
Fiducial model & $2.18\times10^7$ &  $6243$ & $2023$ & $107$ \\
\enddata
\tablecomments{The signs are same as labeled in table \ref{tab:1} and table \ref{tab:2}.
}
\end{deluxetable}

\section{Summary and Conclusion}
\label{sec:6}
With a hybrid BPS model and the Galaxy model, we have modeled the formation and spatial distribution of DDs with ELM WDs in our Galaxy. We have studied the birthrate and semi-detached rate of these systems. In addition, we give a comprehensive discussion about properties (including chirp mass distribution, frequency distribution) of the population of ELM WDs as GW sources. Finally, we present the properties potential candidates from LISA only and LISA+ELM Survey/{\it{Gaia}} observations. Our main conclusions are summarized as follows. 
\begin{itemize}
\setlength{\itemsep}{0pt}
\setlength{\parsep}{0pt}
\setlength{\parskip}{0pt}
	\item[(1)] Most of halo ELM WDs are produced from the RL channel, which is consistent 
      with the observations. They are hardly detected by LISA, due to relatively long orbital periods and a small number.
    \item[(2)] There are $400-6000$ sources having GW signals higher than the LISA sensitivity curve of $\rm {SNR} = 7$, among which about $100-2000$ have chirping signals. 
    \item[(3)] 	{\color{black}In the standard model with $\alpha_{\rm CE}\lambda=1$, about one dozens of LISA detectable sources are possible to be found by ELM Survey or other ground multi-wavelength photometric variability surveys. With the combination of {\it{Gaia}} and LISA there are about $\sim100$ sources are expected to be detected. }
    \item[(4)]  {\color{black}With the combination of EM and GW observations, ELM WDs with mass $\lesssim0.27M_\odot$ have large possibility to be detected, the chirp masses for those systems are generally less than $\sim 0.4M_\odot$, and the orbital period distribution shows the mean value around $\sim22$ min. }
\end{itemize}

\section*{Acknowledgements}
The authors would like to thank Valeriya Korol for providing the table of star formation rates about the BP99 Galactic model. The authors gratefully acknowledge the computing time granted by the 
Yunnan Observatories, and provided on the facilities at the Yunnan 
Observatories Supercomputing Platform. 
This work is partially supported by the Natural Science Foundation of China (Grant no. 
11733008, 11521303, 11703081, 11422324), by the National Ten-thousand talents program, 
by Yunnan province (No. 2017HC018), by Youth Innovation Promotion Association of the 
Chinese Academy of Sciences (Grant no. 2018076) and the CAS light of West China Program. 
\software{BSE \citep{hurley00,hurley02}}
\software{MESA (v9575; \citealt{paxton2011,paxton2013,paxton2015})}

\bibliographystyle{./aasjournal}
\bibliography{./gwr}

\include{table_information}
\include{table_KS}
\include{figure}

\end{document}